\documentclass[%
 aip,
 amsmath,amssymb,
reprint,%
]{revtex4-1}

\usepackage{graphicx}
\usepackage{dcolumn}
\usepackage{bm}

\usepackage[utf8]{inputenc}
\usepackage[T1]{fontenc}
\usepackage{mathptmx}
\usepackage{etoolbox}

\makeatletter
\def\@email#1#2{%
 \endgroup
 \patchcmd{\titleblock@produce}
  {\frontmatter@RRAPformat}
  {\frontmatter@RRAPformat{\produce@RRAP{*#1\href{mailto:#2}{#2}}}\frontmatter@RRAPformat}
{}{}
}%
\makeatother
\begin{document}


\title{Effects of gas--liquid phase transitions on soundwave propagation: A molecular dynamics study}
\author{Yuta Asano}
\affiliation{
    Institute for Solid State Physics, The University of Tokyo, Kashiwa, Chiba 277-8581, Japan
}
\email{yuta.asano@issp.u-tokyo.ac.jp}

\author{Hiroshi Watanabe}%
\affiliation{
    Department of Applied Physics and Physico-Informatics, Keio University, Yokohama, Kanagawa 233-8522, Japan
}%

\author{Hiroshi Noguchi}
\affiliation{%
    Institute for Solid State Physics, The University of Tokyo, Kashiwa, Chiba 277-8581, Japan
}%

\date{\today}

\begin{abstract}
  To understand ultrasonic cavitation, it is imperative to analyze the effects of the gas--liquid phase transitions on soundwave propagation. Since current methods based on fluid dynamics offer limited information, it is imperative to carry out further research on this phenomenon. In this study, we investigated the effects of cavitation and near-critical fluid on soundwaves using the molecular dynamics (MD) simulations of Lennard-Jones fluids. In the first-order liquid-to-gas transition region (far from the critical point), the waveform does not continuously change with the temperature and source oscillation amplitude owing to the discontinuous change in the density due to the phase transition. Meanwhile, in the continuous transition region (crossing near the critical point), the waveform continuously varies with temperature regardless of the amplitudes because phase separation is not involved in this region. The density fluctuations increase as the amplitude increases; however, it does not affect the waveform. Thus, we clarified that the first-order and continuous transitions have different impacts on soundwaves. Moreover, we determined the acoustic characteristics, such as attenuation and nonlinear parameters, by comparing the results of the numerical solution of Burgers' equation and MD simulation. Burgers' equation clearly describes the soundwave phenomenon until phase separation or bubble formation occurs. In the continuous transition region, the attenuation parameters tend to diverge, reflecting a critical anomaly trend. We observed the bubbles move forward with the oscillation of their radii owing to their interaction with the soundwaves. This is the first direct observation of the interaction using MD simulations.
\end{abstract}

\maketitle

\section{Introduction}
Ultrasonic waves are soundwaves with frequencies above $20$ kHz, which cannot be heard by the human ear. Ultrasonic waves are used in various fields owing to their ability to propagate through a wide range of solid, liquid, and gaseous media. For example, metal processing uses the energy of ultrasound, and sonar uses its propagation properties to detect objects. In energy applications, ultrasonic cavitation is used in chemical reactions, wastewater treatment, medical treatment, food processing, and casting among others~\cite{td99, lkg07,  kf14, ws15, mason16, etw19, ibc19, at20, savun20,  soyama20, ypl20}.
Ultrasonic cavitation is induced by irradiating a liquid with powerful ultrasonic waves for bubble generation. The bubbles expand, deflate, and occasionally collapse because of their interaction with the soundwaves. A high-temperature and high-pressure field is locally generated by the bubble collapse. In addition, a jet is also generated when the bubble collapse occurs near the wall of an object~\cite{wc97}. These chemical and physical effects cause molecular dissociation and peeling off of the adhered materials. Ultrasonic cavitation treatment has potential applications in various fields as an earth-conscious technology because of its low environmental impact due to the localized nature of the extreme environment. Therefore, a thorough understanding of the mechanism of ultrasonic cavitation is vital for fluid engineering and its application in various fields. However, ultrasonic cavitation is a highly complex phenomenon that involves the repeated generation and dissipation of bubbles under highly nonequilibrium conditions. Thus, several aspects of this phenomenon remain unclear.

Ultrasonic horns are widely used in laboratory experiments to generate ultrasonic cavitation, such as for the preparation of dispersions.  To date, several experimental studies have been conducted on ultrasonic cavitation using ultrasonic horns~\cite{mgd03, dvc10, zmc14, tle17, fyk18, ysp21}. These experiments mainly focused on observing the dynamics of the bubbles using video cameras and measuring the sound pressure using hydrophones. As cavitation occurs, numerous bubbles are generated below the horn, which form a conical bubble structure~\cite{mgd03}. The interior of a bubble structure is known to have high chemical activity~\cite{dvc10}. In addition, the sound pressure signal during cavitation is characterized by the generation of harmonics due to nonlinearity, subharmonics of large bubble clusters attached to the horn, and broadband signals from the collapse of bubbles with various sizes~\cite{hcz04}. The cavitation behavior with respect to different parameters, such as horn shape, diameter, frequency, amplitude, and physical properties of the liquid has been investigated~\cite{zmc14, tle17, fyk18, ysp21}. For industrial applications, it is crucial to understand the cavitation morphology and sound pressure characteristics for scaling up the device. However, a major problem is that cavitation only occurs in a limited area near the horn because the bubbles prevent soundwave propagation, which is essential for cavitation inception. Therefore, studies have been conducted to efficiently utilize cavitation, such as placing a channel close to the horn~\cite{ty19}. Moreover, it is essential to understand the interactions between soundwaves and bubbles for appropriate cavitation control.

The acoustic properties of soundwaves propagating in a fluid, such as speed of sound, attenuation, and nonlinearity, are evaluated based on the physical properties of the fluid, including elastic modulus and viscosity. For a liquid with bubbles, these properties are significantly altered depending on the bubble size and content~\cite{silberman57, wrc05}. Thus, deeper understanding of ultrasonic cavitation could be obtained by establishing a theoretical model for soundwaves propagating in bubbly liquids~\cite{kyw10, louisnard12, fcc14, prosperetti15, zgd18, zgg18, trujillo18}, which has also been applied to ultrasonic cavitation by ultrasonic horns~\cite{trujillo20}. These theories are based on fluid dynamics, where a small number of parameters, such as bubble size distribution and content, describes the bubbles. For a more accurate description of the sound field, it is important to consider the effects of bubble formation, growth, splitting, coalescence, and collapse. Further, it is necessary to understand the interactions between the liquid and bubbles, such as bubble--liquid and bubble--bubble interaction; the former is related to the first Bjerknes force, whereas the latter is related to the second Bjerknes force~\cite{crum75,louisnard08}. Researches on these interactions have been conducted mainly through computer simulations~\cite{yit08, fm15, xu18, mhl18, qmh18, bb19, pandey19, yk20}.  However, despite the gas--liquid phase transition as the origin of bubble inception, it is challenging to analyze the phase transition effects on the soundwaves based on fluid dynamics.

In this study, we analyzed soundwave propagation in fluids involving gas--liquid phase transitions through molecular dynamics (MD) simulations. Although analyses of bubble vibration, and multiple bubble formation and growth in a stationary liquid have already been conducted~\cite{oi03, hlg10, wih12}, this study is the first attempt to analyze ultrasonic cavitation induced by soundwaves on a molecular scale. In our previous study~\cite{awn20b}, sound propagation in a fluid in the absence of phase transitions was quantitatively analyzed using MD simulations. The K\'arm\'an vortex~\cite{awn18,awn19}, including cavitation~\cite{awn20a,awn21}, has also been simulated by MD. Here, we investigate the effect on soundwaves for the first-order transition region from the liquid to gas phases, which are observed far from the critical point, and continuous transition region near the critical point. We also estimated the acoustic parameters, such as attenuation coefficient and nonlinearity, based on the hydrodynamic calculations. Finally, we performed large-scale MD simulations to investigate the effects of bubble generation and oscillation on the soundwaves.

The rest of the paper is organized as follows: Section II describes the simulation model and method. Section II.B describes the protocol of acoustic parameter estimation. The results are presented in Sec. III. Section IV presents a summary and discussion.
\section{Method}
\subsection{MD simulation}
In this study, the fluid is modeled to consist of monatomic molecules with a smoothed-cutoff Lennard-Jones (LJ) potential interaction:
\begin{eqnarray}
    u_{\rm sLJ}(r;r_{\rm c})&=&
    \left\{
    \begin{array}{ll}
        u_{\rm LJ}(r) - u_{\rm LJ}(r_{\rm c}) - (r-r_{\rm c})u'_{\rm LJ}(r_{\rm c}) & (r\le r_{\rm c}) \\
        0                                                                           & (r>r_{\rm c})
    \end{array} \right.,\label{eq:sLJ}\\
    u_{\rm LJ}(r)&=&4\epsilon\left[\left(\frac{\sigma}{r}\right)^{12} - \left(\frac{\sigma}{r}\right)^{6}\right],
    \label{eq:LJ}
\end{eqnarray}
where $r$ is the intermolecular distance, and $r_{\rm c}$ is the cutoff distance of the potential function, which is set to $r_{\rm c}=2.5\sigma$. The prime symbol ($'$) in Eq.~(\ref{eq:sLJ}) represents the derivative with respect to $r$. Hereafter, all physical quantities are expressed in units of length $\sigma$, energy $\epsilon$, and time $\tau=\sqrt{m\sigma^2/\epsilon}$ ($m$: mass of a molecule).

\begin{figure}
    \includegraphics[width=3.37in]{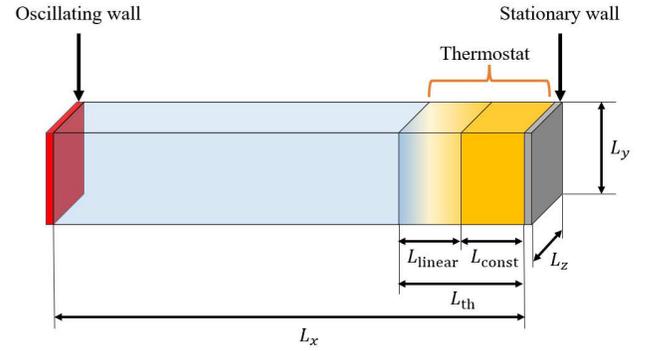}%
    \caption{\label{fig:1} Schematic of the computational domain. The positions of the oscillating and stationary walls are $x=0$ (red) and $x=L_x$ (gray), respectively. Two orange boxes on the stationary wall indicate the equilibration area by the Langevin thermostat. The gradation of the orange color in the left box schematically represents a linear increase in the friction constant of the thermostat.}%
\end{figure}

The simulation box is a rectangular parallelepiped with dimensions $L_x\times L_y\times L_z =50~000\times 25 \times 25$ (Fig.~\ref{fig:1}). Periodic boundary conditions are applied along the $y$- and $z$-axes. The oscillating and stationary walls are placed at $x=0$ and $50~000$, respectively. These walls are modeled by fixing the LJ molecules on the square lattice points of the $yz$-plane, where the surface density of the wall molecules is $0.82$. The interactions between the wall and fluid molecules are $u_{\rm sLJ}(r;2.5)$. The orange and gradation areas ($49~000\le x \le 50~000$, i.e., $L_{\rm th}=1000$) in Fig.~\ref{fig:1} impose the Langevin thermostat to annihilate soundwaves~\cite{awn20b}. The friction coefficient of the Langevin thermostat linearly increases with $x$ from $0.0001$ to $0.1$ in $49~000 \le x < 49~500$ (gradation color in Fig.~\ref{fig:1}, $L_{\rm linear}=500$) and $0.1$ in the rest of the area (orange color in Fig.~\ref{fig:1}, $L_{\rm const}=500$).

\begin{figure}
    \includegraphics[width=3.37in]{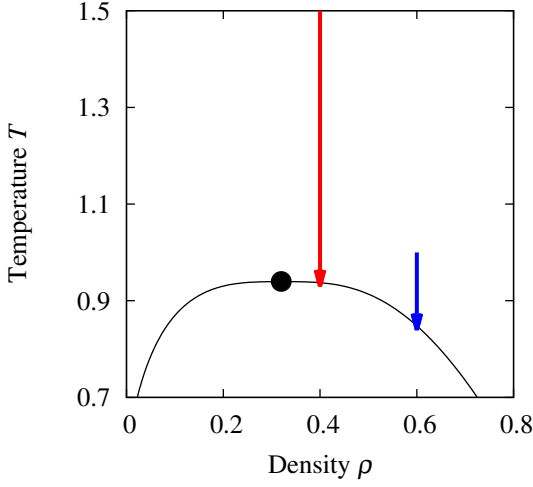}%
    \caption{\label{fig:2} Thermodynamic state of the LJ fluid in the simulation. The input temperature $T_0$ is varied with a constant mean density $\rho_0=0.6$ and $0.4$ along the blue and red arrows, which represent the first-order transition region and continuous transition region, respectively. The black line denotes the liquid--gas phase boundary of the LJ system, and the black circle is the critical point ($T_{\rm c}=0.94$, $\rho_{\rm c}=0.32$)~\cite{awn20a}.}%
\end{figure}

At the initial time, the fluid molecules are randomly placed without overlapping in the simulation box, except at the wall position. We study two densities, $\rho_{0} = 0.6$ and $0.4$ with $18~749~626$ and $12~499~751$ molecules, respectively. The former examines the soundwaves in the first-order transition region far from the critical point, and the latter in the continuous transition region crossing near the critical point, respectively. The temperature ranges of the two systems are shown in Fig.~\ref{fig:2}. The gas--liquid phase boundary of the LJ system (the black line in Fig.~\ref{fig:2}) was obtained by the gas--liquid coexistence simulation~\cite{awn20a}. The initial velocities of the fluid molecules are generated according to the Maxwell distribution with temperature $T_{0}$ and zero average velocity. The wall, which is the sound source, oscillates sinusoidally in the $x$-direction. The position of the oscillating wall at time $t$ is given by:
\begin{eqnarray}
    x_{\rm w}(t)=A \sin \left(2\pi f t\right),
\end{eqnarray}
where  $A$ and $f$ are the oscillation amplitude and frequency, respectively.
A low frequency $f=0.001$ is chosen for minimal influence of the nonlinear effects, as indicated in a previous study~\cite{awn20b}.

LAMMPS (Large-scale Atomic/Molecular Massively Parallel Simulator)~\cite{plimpton95} is used for the numerical integration of the equations of motion of the molecules. The velocity Verlet algorithm is adopted for the numerical integration. The time increment is $\Delta t= 0.004$, and the maximum number of steps is approximately $20~000~000$. A total of $1~000~000$ steps is needed for the initial equilibration, and $5~000~000$ steps are required to reach a steady wave propagation. The remaining steps are used to estimate the sound field. The statistical averages and errors are evaluated using three or more independent simulations.

The waveform is obtained from the spatial variation of the mean flow velocity $\bar{V}^{x}$ in the $x$-direction. The simulation box is divided into slices with a thickness of $\Delta L_x=25$. For each slice, the instantaneous flow velocity at time $t$ is obtained as:
\begin{eqnarray}
    V_{i}^{x}(t)&=&\frac{1}{n_{i}(t)}\sum_{k \in G_i}v_{k}^{x}(t),\label{eq:v}
\end{eqnarray}
where $n_{i}(t)$ is the number of molecules in the $i$th slice at time $t$, $v_{k}^{x}(t)$ is the $x$-component of the velocity of the $k$th molecule at time $t$, and $G_i$ is the set of molecules in the $i$th slice. Here, a time-synchronous averaging is used to obtain the waveform $\bar{V}^{x}$:
\begin{eqnarray}
    \bar{V}^{x}_{i}(\tau_j)&=&\frac{1}{M_{\rm osc}}\sum_{k=0}^{M_{\rm osc}-1}V^{x}_{i}\left(\tau_j+\frac{k}{f}\right),
\end{eqnarray}
where $M_{\rm osc}$ is the number of oscillations in the oscillating wall during the observation, $\tau_{j}=j/(fN_{\rm p})~(j=0,1,\cdots,N_{\rm p}-1)$, which is the phase of the oscillating wall, and $N_{\rm p}$ is the number of divisions in phase, which is set to $N_{\rm p} = 200$. The density profile is calculated using the same slice: $\rho_i(t)=n_i(t)/V_{\rm slice}$, where the slice volume $V_{\rm slice} = 15~625$.

Furthermore, the density field is estimated by dividing the simulation box into cells with the dimensions $l_x\times l_y\times l_z = 2.5\times 2.5\times 2.5$.

\subsection{Acoustic properties}
Hydrodynamic calculations are performed to investigate the influence of phase transitions on the acoustic properties. The typical parameters considered in this study are the speed of sound, attenuation coefficient, and nonlinearity parameter. These parameters are estimated using Burgers' equation, as follows~\cite{burgers48}:
\begin{eqnarray}
    \frac{\partial p_{\rm a}}{\partial x} + \frac{1}{c_{0}}\frac{\partial p_{\rm a}}{\partial t} - \frac{b}{2c_{0}}\frac{\partial^2 p_{\rm a}}{\partial t^2} - \frac{\beta p_{\rm a}}{\rho_{0}c_{0}^3}\frac{\partial p_{\rm a}}{\partial t} = 0,
\end{eqnarray}
where $p_{\rm a}$, $c_{0}$, $b$, and $\beta$ are the fluctuating component of the pressure, speed of sound, attenuation parameter, and nonlinear parameter, respectively. We search for a set of parameters ($c_{0}$, $b$, $\beta$) to ensure that the numerical solution of Burgers' equation is compatible with the waveform obtained from the MD simulation. Thus, the parameters are explored that minimize the loss function, $\chi^2$, defined as follows:
\begin{eqnarray}
    \chi^2=\frac{1}{M_{\rm osc}N_{\rm p}}\sum_{i, j} \left(\frac{\bar{V}^{x}_{i}(\tau_j)}{V_{\rm w}^{\rm max}} - \frac{p_{\rm a}(x_{i},\tau_j)}{p_{\rm a}^{\rm max}}\right)^2, \label{eq:chi2}
\end{eqnarray}
where $V_{\rm w}^{\rm max}$ is the maximum velocity of the oscillating wall. The plane wave relation $p_{\rm a}=\rho_{0} c_{0} \bar{V}^{x}$ is assumed between the velocity and pressure fluctuation components. The maximum value of the pressure fluctuation is $p_{\rm a}^{\rm max}=\rho_{0} c_{0} V_{\rm w}^{\rm max}$. Burgers' equation is discretized and numerically integrated using the central difference method in the time direction and second-order backward difference method in the spatial direction~\cite{mohamed19}. The oscillation condition of the wall at $x=0$ and periodic boundary condition along the time direction are imposed as follows:
\begin{eqnarray}
    p_{\rm a}(x=0, t)&=& p_{\rm a}^{\rm max}\sin\left(2\pi f t\right),\\
    p_{\rm a}(x, t)&=&p_{\rm a}\left(x, t+\frac{1}{f}\right).
\end{eqnarray}
The widths of the discretization in time and space are $\Delta t = 0.5$ and $\Delta x = 3.125$, respectively.

\section{Results}
\subsection{First-order transition region}
\begin{figure}
    \includegraphics[width=3.37in]{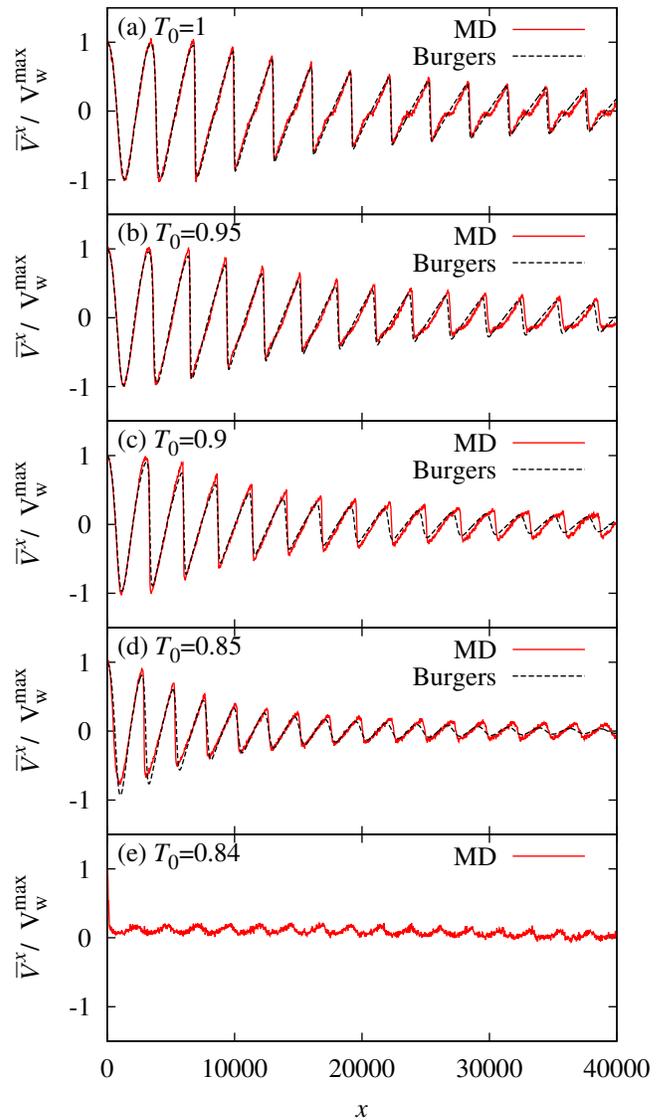}%
    \caption{\label{fig:3} Temperature $T_{0}$ dependence of the waveform in the first-order transition region (density $\rho_{0}=0.6$). (a) $T_{0}=1$, (b) $T_{0}=0.95$, (c) $T_{0}=0.9$, (d) $T_{0}=0.85$, and (e) $T_{0}=0.84$. The boiling temperature is $T_{\rm b}=0.85$, and the amplitude of the oscillating wall is $A=10$. As the temperature decreases, the wavelength decreases, and the attenuation rate of the soundwave increases. Although the change until $T_{0}=0.85$ is continuous, that of $T_{0}=0.84$ is not continuous. The rapid change is attributed to the phase transition. The flow velocity is normalized by the maximum velocity $V_{\rm w}^{\rm max}$ of the oscillating wall, which was also applied to Figs.~\ref{fig:4}, \ref{fig:7}, \ref{fig:8}, and \ref{fig:12}. The black dashed lines indicate the numerical solution of Burgers' equation by parameter fitting.
    }%
\end{figure}

\begin{figure}
    \includegraphics[width=3.37in]{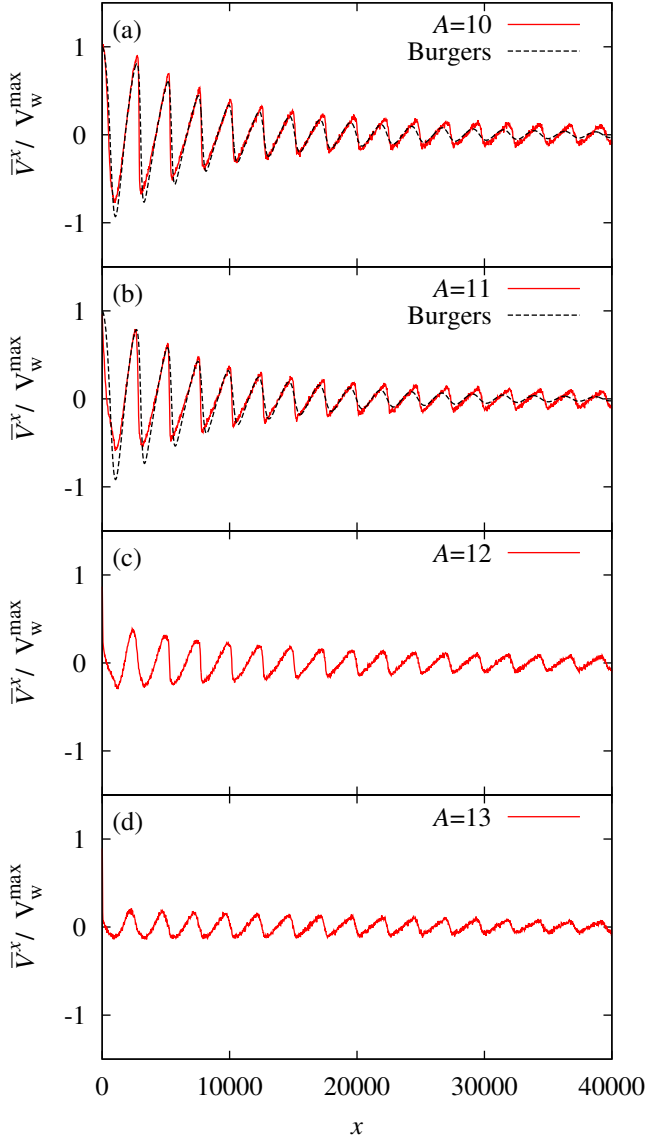}%
    \caption{\label{fig:4} Dependence of the waveform on the amplitude $A$ of the oscillating wall at the first-order transition point ($\rho_{0}=0.6$, $T_{0}=0.85$). (a) $A=10$, (b) $A=11$, (c) $A=12$, and (d) $A=13$. The black dashed lines show the numerical solution of Burgers' equation obtained by parameter fitting.}%
\end{figure}

Figure~\ref{fig:3} shows the temperature dependence of the waveform in the first-order transition region ($\rho_{0}=0.6$). The waveform is normalized using the maximum velocity $V_{\rm w}^{\rm max}$ of the oscillating wall, which is also applied to the waveforms of the subsequent figures. As the temperature decreases from $T_{0}=1$ to $T_{0}=0.85$ (boiling point), the wavelength becomes shorter, and the attenuation rate increases. In contrast, at $T_{0}=0.84$, the wave rapidly decays near the oscillating wall. After the decay, a small-amplitude wave propagates. Figure~\ref{fig:4} shows the amplitude $A$ dependence of the waveform at the boiling point ($\rho_{0}=0.6$, $T_{0}=0.85$). In the amplitude range of $A \le 11$, there are no considerable changes in the waveforms. In contrast, at $A \gtrsim 12$, the wave significantly decays near the oscillating wall. In addition, a small-amplitude wave propagates after the decay. This sudden change is similar to that at a low temperature, which resulted in a smaller amplitude, as shown in Fig.~\ref{fig:3}(e).


\begin{figure}
    \includegraphics[width=2.7in]{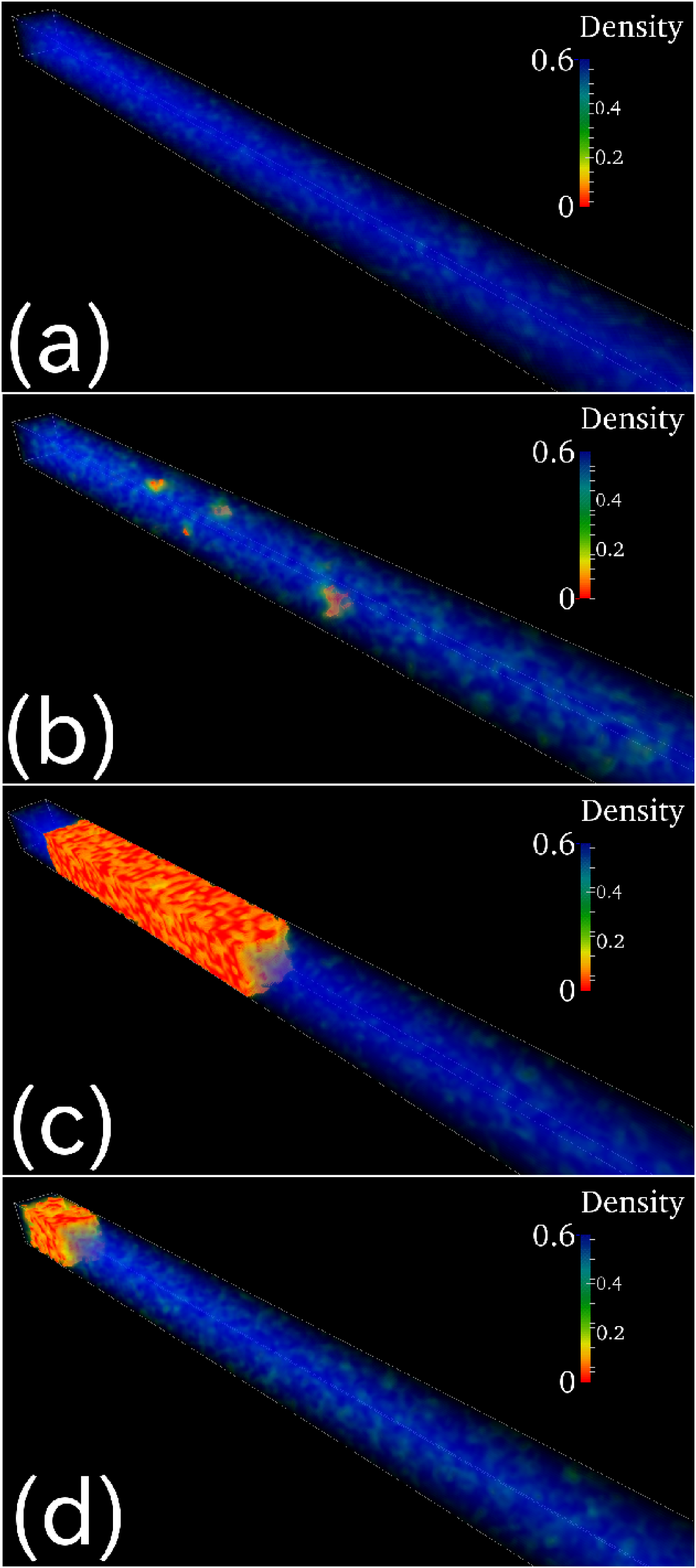}%
    \caption{\label{fig:5} Typical snapshots of the density field in the first-order transition region ($\rho_{0}=0.6$). The liquid and gas phases are displayed in blue and red, respectively.
(a) The liquid phase occupies the entire area at the temperature $T_{0}=1$ and oscillation amplitude $A=10$.
(b)  A few gas bubbles appear at the boiling point $T_{0}=0.85$ and $A=10$.
(c) $T_{0}=0.84$ and $A=10$. (d) $T_{0}=0.85$ and $A=13$.
A further decreasing temperature or a strengthening of the oscillation generates a phase separation.}%
\end{figure}

If the sudden changes in the waveforms shown in Figs.~\ref{fig:3} and \ref{fig:4} are attributed to the phase transition, the influence should be visible in the density field. Figure~\ref{fig:5} shows the typical density field, where the blue and red cells indicate the liquid and gas phases, respectively. Figures~\ref{fig:5}(a), (b), and (c) show the density changes as the temperature $T_{0}$ decreases under a constant amplitude, $A=10$. The corresponding movies of Figs.~\ref{fig:5}(b) and (c) are shown in Movies 1 and 2 in the supplementary material, respectively. At a high temperature ($T_{0}=1$), the entire area is in the liquid state. At the boiling point, the bubble nuclei appear in some portions and subsequently disappear (Fig.~\ref{fig:5}(b)). In the phase separation region, the gas phase appears in the vicinity of the oscillating wall (Fig.~\ref{fig:5}(c)). Meanwhile,  the liquid phase exists between the oscillating wall and gas phase. Figure~\ref{fig:5}(d) shows the density field for a larger amplitude ($A=13$) at the boiling point, which corresponds to the condition in Fig.~\ref{fig:4}(d). Although the gas phase appears, the location and extent of the gas phase differ from that in Fig.~\ref{fig:5}(c). At $A=12$, as shown in Fig.~\ref{fig:4}(c), phase separation occurs in one run, but not in the other two runs. Thus, this amplitude is close to the threshold required to induce separation. Therefore, the average waveform is an intermediate between that in $A=11$ and $A=13$. Thus, in the first-order transition region, the density field changes discontinuously due to the phase transition in response to the changes in temperature and amplitude, resulting in sudden changes in the waveform. Figure~\ref{fig:4} shows that the phase transition does not affect the wavelength, which implies that the gas phase does not affect the speed of sound.

\begin{figure}
    \includegraphics[width=3.37in]{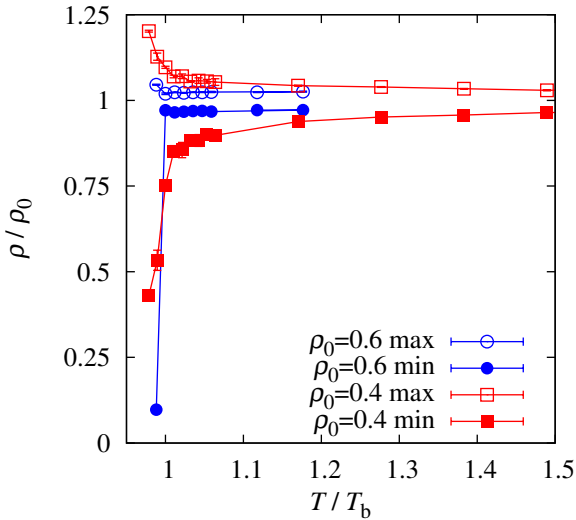}%
    \caption{\label{fig:6} Temperature $T_{0}$ dependence of the maximum and minimum values of the time-averaged density profile $\rho$ in the system at the oscillation amplitude $A=10$. The density and temperature are normalized by the initial density $\rho_{0}$ and boiling temperature $T_{\rm b}$, respectively. For the first-order transition region (blue), the minimum density shows a sharp drop in the vicinity of the boiling point. In contrast, the density changes continuously in the continuous transition region (red). The behavior of the change in density fields reflects the characteristics of the phase transition.}%
\end{figure}

The variations in the density at each slice reveal the discontinuous change in the density field. Figure~\ref{fig:6} shows the temperature dependence of the maximum and minimum values of the time-averaged density profile. The minimum value of the density exhibits a discontinuous change at the boiling point. Therefore, the density field varies discontinuously owing to the gas phase caused by the phase transition, thereby changing the waveform.

\subsection{Continuous transition region}
\begin{figure*}
    \includegraphics[width=6.69in]{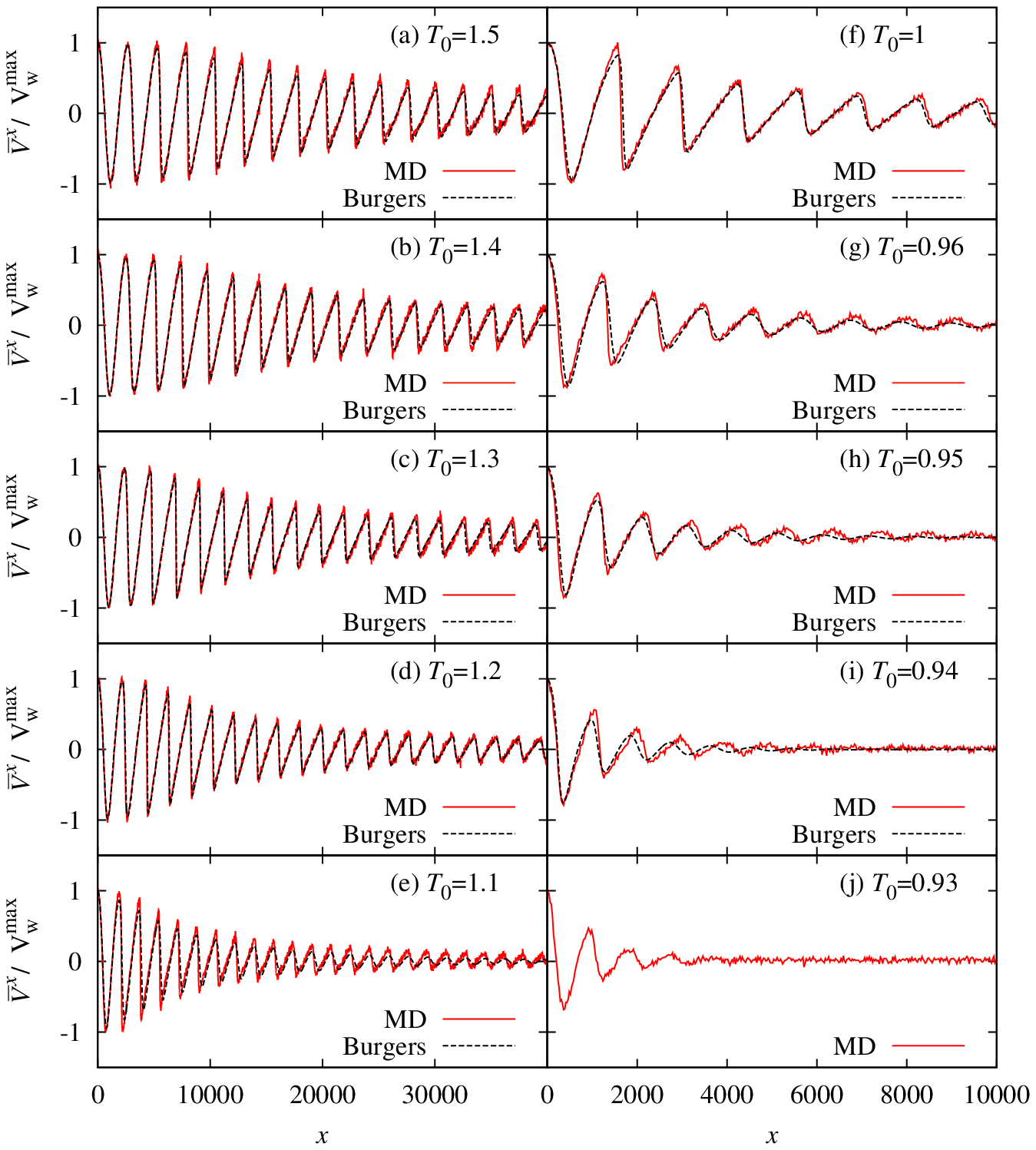}%
    \caption{\label{fig:7} Temperature $T_{0}$ dependence of the waveform in the continuous transition region ($\rho_{0}=0.4$)
at the oscillation amplitude $A=10$. (a) $T_{0}=1.5$, (b) $T_{0}=1.4$, (c) $T_{0}=1.3$, (d) $T_{0}=1.2$, (e) $T_{0}=1.1$, (f) $T_{0}=1$, (g) $T_{0}=0.96$, (h) $T_{0}=0.95$, (i) $T_{0}=0.94$, and (j) $T_{0}=0.93$. The boiling temperature is $T_{\rm b}=0.94$.  As the temperature decreases, the wavelength decreases, and the attenuation rate increases. Moreover, the rate of change increases with approaching the boiling point. The waveform changes continuously against the temperature, even in the phase-separation region. Black dashed lines show the numerical solution of Burgers' equation by the parameter fitting.}%
\end{figure*}

\begin{figure}
    \includegraphics[width=3.37in]{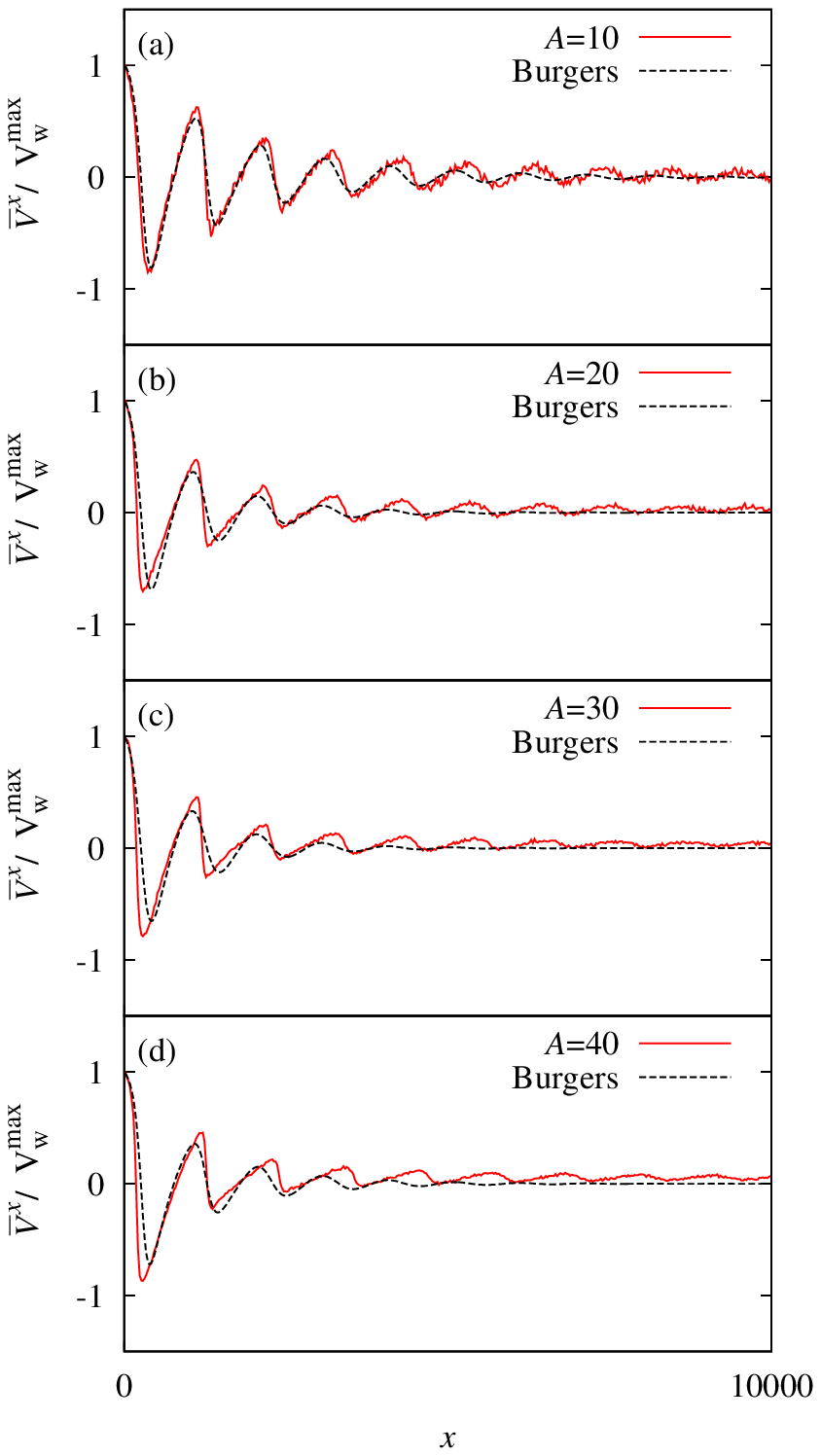}%
    \caption{\label{fig:8} Dependence of the waveform near the boiling point ($\rho_{0}=0.4$, $T_{0}=0.95$) in the continuous transition region on the amplitude $A$ of the oscillating wall. (a) The amplitude $A=10$, (b) $A=20$, (c) $A=30$, and (d) $A=40$. The waveform is independent of the amplitude, and no discontinuous changes are observed. Black dashed lines show the numerical solution of Burgers' equation by the parameter fitting.}%
\end{figure}

Figures~\ref{fig:7} and \ref{fig:8} show the waveform dependence on  the temperature $T_{0}$ and oscillation amplitude $A$ in the continuous transition region ($\rho_{0}=0.4$), respectively. With decreasing temperature, the wavelength decreases, and the amplitude attenuation rate increases. At $T_{0}<1$, the rate of change increases. With increasing amplitude $A$, the amplitude of the velocity oscillation increases linearly, while the waveform remains unchanged. Unlike in the first-order transition region, the waveform does not exhibit discontinuous changes at the boiling point $T_{\rm b}=0.94$, regardless of the variations of $T_{0}$ or $A$.

\begin{figure*}
    \includegraphics[width=6.69in]{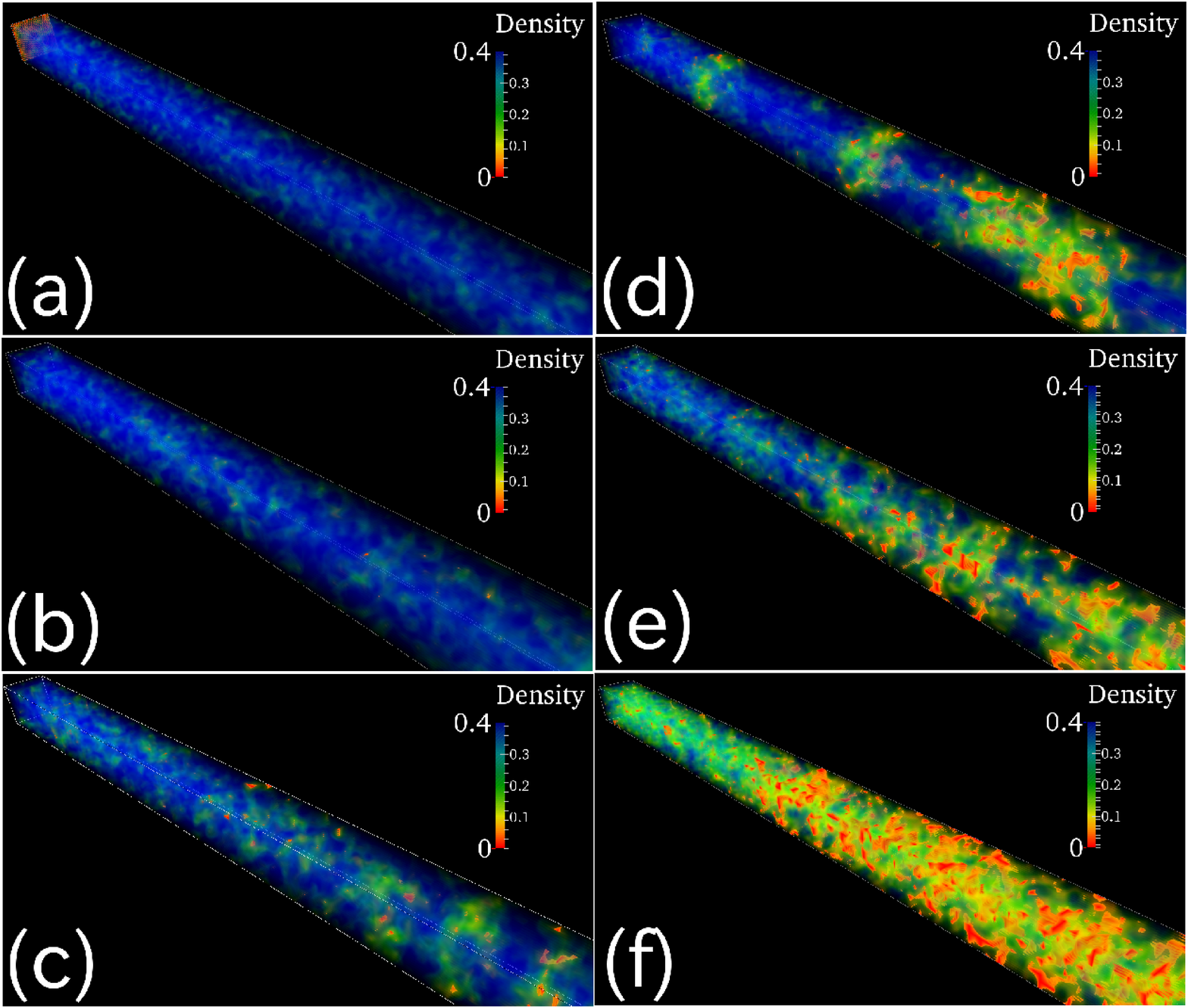}%
    \caption{\label{fig:9} Typical snapshots of the density field in the continuous transition region ($\rho_{0}=0.4$). 
(a) The temperature $T_{0}=1.2$ and oscillation amplitude $A=10$. (b) $T_{0}=1$ and $A=10$. (c) $T_{0}=0.95$ and $A=10$. (d) $T_{0}=0.93$ and $A=10$. (e) $T_{0}=0.95$ and $A=20$. (f) $T_{0}=0.95$ and $A=40$. 
Low-density parts appear as the temperature decreases. In the phase-separation region 
($T_0<T_{\rm b}=0.94$), a slight tendency of phase separation is observed. 
However, no clear phase separation occurs. The density fluctuations increase with increasing amplitude.}%
\end{figure*}

Figure~\ref{fig:9} shows the density fields in the continuous transition region. Figures~\ref{fig:9}(a)--(d) show the density field with decreasing temperature at a constant amplitude (Movie 3 in the supplementary material, corresponding to Fig.~\ref{fig:9}(d)). At a high temperature ($T_{0}=1.2$), the density is approximately homogeneous. Moreover, the low-density area gradually increases with decreasing temperature. At $T_{0}=0.93$, the density field exhibits a slight tendency toward phase separation. However, this separation is not prominent, unlike that of the first-order transition region. As shown in Fig.~\ref{fig:6}, the density fluctuation continuously increases with decreasing temperature in the continuous transition region. Therefore, the waveform continuously varies with the temperature change. In contrast, increasing the amplitude at the boiling point amplifies the overall density fluctuation (Figs.~\ref{fig:9}(e) and (f), and Movie 4, corresponding to Fig.~\ref{fig:9}(f), in the supplementary material), whereas the waveform remains unchanged. Furthermore, in contrast to the first-order transition region, phase separation does not occur. Thus, the behavior of the density field, and consequently, its effects on the soundwave are different for the first-order and continuous transition regions.

\subsection{Acoustic properties}
\begin{figure}
    \includegraphics[width=3.37in]{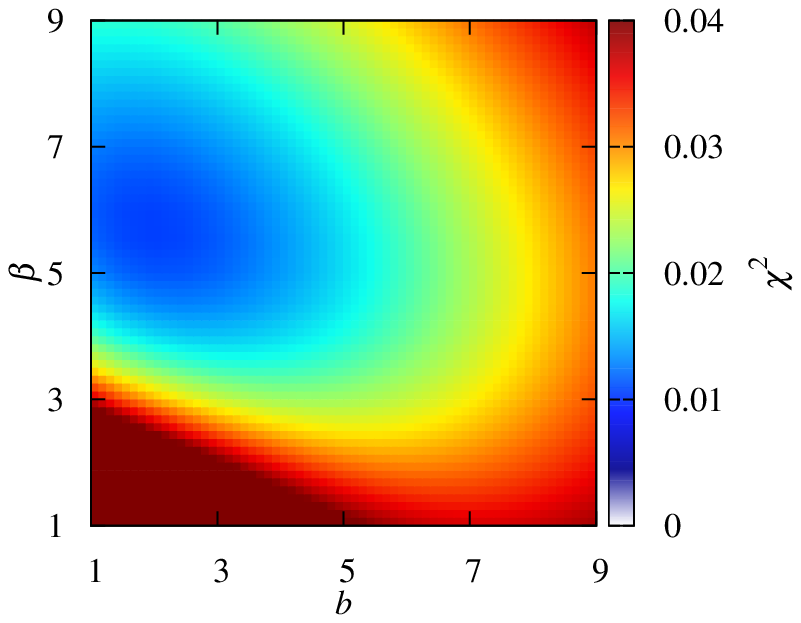}%
    \caption{\label{fig:10} Distribution of the loss function $\chi^2$ at temperature $T_{0}=0.95$ and density $\rho_{0}=0.6$ in the first-order transition region at the oscillation amplitude $A=10$. $\chi^2$ reaches a minimum at the attenuation parameter $b=2.0$ and the nonlinear parameter $\beta=5.5$.}%
\end{figure}

We determined the parameters of Burgers' equation to clarify the effects of the phase transition. As shown in Figs.~\ref{fig:4} and \ref{fig:8}, the wavelength is not affected by the phase transition. We use the adiabatic speed of sound $c_{S}$ obtained by equilibrium MD simulation~\cite{awn20b} for $c_{0}$. Therefore, the fitting parameters are the attenuation parameter $b$ and nonlinear parameter $\beta$. Figure~\ref{fig:10} shows the results of the parameter search in the first-order transition region. The minimum loss function $\chi^2$ is obtained at $b=2.0$ and $\beta=5.5$. The intervals for $b$ and $\beta$ are $\Delta b = 0.5$ and $\Delta \beta = 0.5$, respectively. Similar parameters searches yield the parameters for all waveforms of the MD simulations, except for the phase-separated state, as shown in Figs.~\ref{fig:3}(e) and \ref{fig:4}(d). The numerical solutions of Burgers' equation are indicated by the black lines in Figs.~\ref{fig:3}, \ref{fig:4}, \ref{fig:7}, and \ref{fig:8}. Overall, Burgers' equation reproduces the results of the MD simulation well with only slight deviations at a distance from the oscillating wall due to thermal fluctuations.

\begin{figure}
    \includegraphics[width=3.37in]{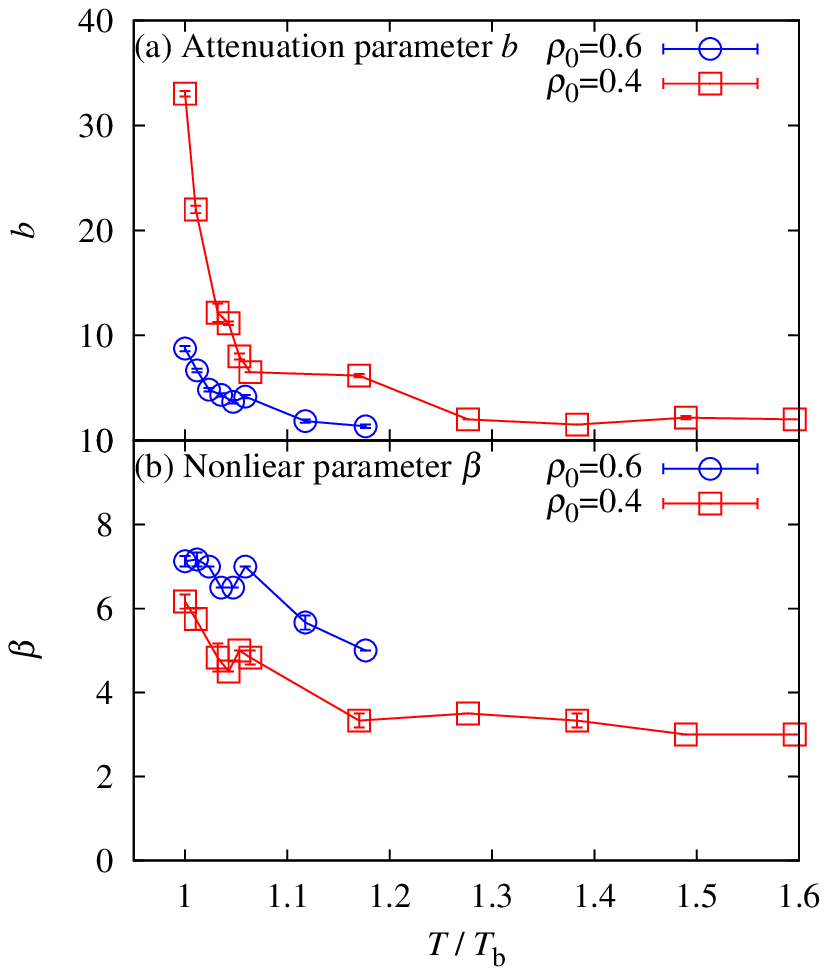}%
    \caption{\label{fig:11} Temperature dependence of (a) attenuation parameter $b$ and (b) nonlinear parameter $\beta$ at the oscillation amplitude $A=10$. The temperature $T_{0}$ is normalized by the boiling temperature $T_{\rm b}$.
    }%
\end{figure}

Figure~\ref{fig:11} shows the temperature dependence of the attenuation and nonlinear parameters obtained by fitting. In both the first-order and continuous transition regions, the attenuation parameter increases with decreasing temperature. For the continuous transition regions, the attenuation parameters exhibit divergent behavior, reflecting the effects of the critical anomalies. In the first-order transition region, the attenuation parameter exhibits only a small change near the transition point. The nonlinear parameters also exhibit an increasing linear trend with decreasing temperature. The attenuation is predominant in the continuous transition region, whereas nonlinearity is more dominant in the first-order transition region. Thus, Burgers' equation allows the estimation of the acoustic properties.

\subsection{Bubble effects}
In the first-order transition region, a layer of the gas phase is formed parallel to the wall due to the interconnection of the bubbles through the periodic boundary. Hence, a larger system size is required to obtain spherical bubbles with long life. Therefore, we expand the dimensions of the simulation box to $L_x \times L_y \times L_z =10~000 \times 100 \times 100$. The Langevin thermostat area is $9000 \le x \le 10~000$ ($L_{\rm th}=1000$), where the friction coefficient increases linearly at $9000\le x < 9500$ ($L_{\rm linear}=500$) and is equal to a constant value of $0.1$ at $9500\le x \le 10~000$ ($L_{\rm const}=500$, Fig.~\ref{fig:1}). The thermodynamic state is set on the first-order transition point ($\rho_{0}=0.6$ and $T_{0}=0.85$) with $59~994~001$ molecules. The computations are executed on $144$ nodes of the ISSP supercomputer (AMD EPYC 7702, $64$ cores $\times 2$ per node). The simulation is parallelized using the flat message passing interface (MPI) ($18~432$ MPI processes), and its execution time for $15~000~000$ steps (time $t=60~000$ in LJ unit) is approximately $50$ h.

\begin{figure}
    \includegraphics[width=3.37in]{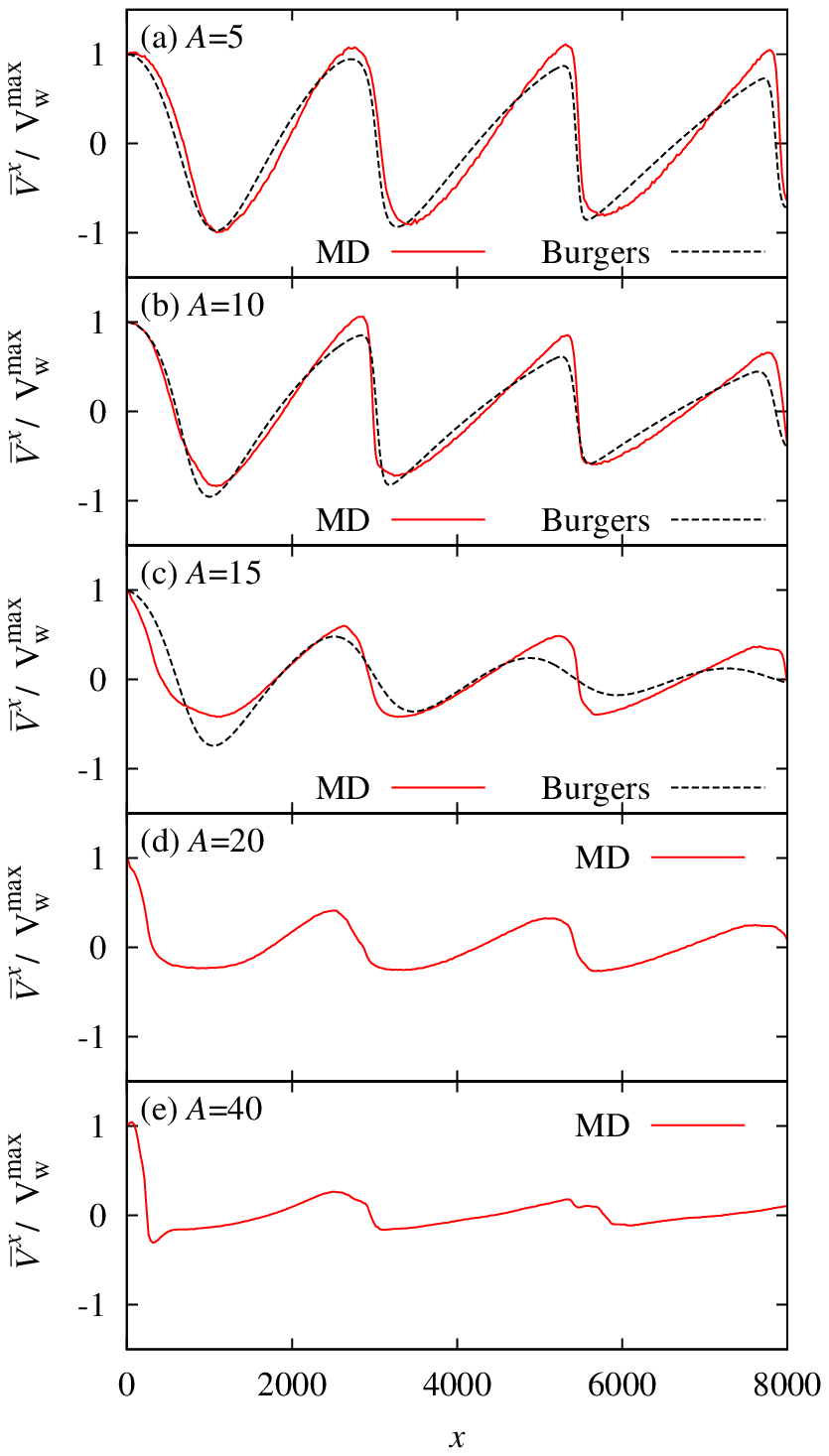}%
    \caption{\label{fig:12} Dependence of the waveform at the first-order transition point ($\rho_{0}=0.6$, $T_{0}=0.85$) on the amplitude $A$ of the oscillating wall for large-scale simulations. (a) $A=5$, (b) $A=10$, (c) $A=15$, (d) $A=20$, and (e) $A=40$.  The waveform changes are almost negligible in the range of $5\le A \le 10$. However, for $A\ge 15$, the attenuation rate of the wave increases with increasing amplitude $A$. Moreover, the shape of the wave changes. Although such a change occurs, the wavelength remains unchanged (approximately three waves exist in all conditions). Black dashed lines show the numerical solution of Burgers' equation by the parameter fitting.}%
\end{figure}

Figure~\ref{fig:12} shows the amplitude $A$ dependence of the waveform. As shown in Fig.~\ref{fig:4}, there are no changes in the waveform for small amplitudes (Figs.~\ref{fig:12}(a) and (b)). At $A=15$, the attenuation rate of the soundwave slightly increases. The sudden change in the waveform has the same tendency as that of small systems. However, the waveform changes are considerably gradual compared to the phase separation, as shown in Fig.~\ref{fig:4}. The attenuation rate of the soundwave increases with a further increase in the amplitude. In contrast to the results for the small system, the attenuation rate exhibit a continuous increase.

\begin{figure}
    \includegraphics[width=2.0in]{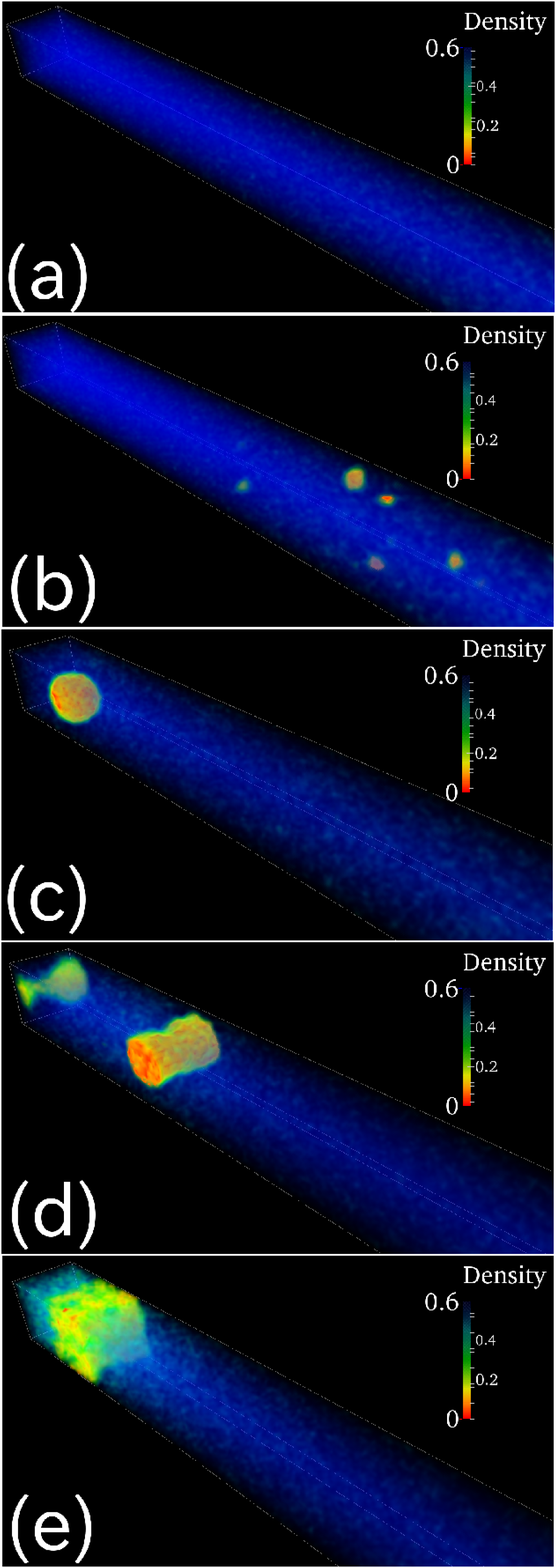}%
    \caption{\label{fig:13} Typical snapshots of the density field in the first-order transition point ($\rho_{0}=0.6$, $T_{0}=0.85$). (a) The oscillation amplitude $A=5$, (b) $A=10$, (c) $A=15$, (d) $A=20$, and (e) $A=40$. The liquid and gas phases are displayed in blue and red, respectively. At $A=5$, the entire area is in the liquid phase. At $A=10$, small bubbles appear instantaneously in various locations and immediately disappear. When the amplitude reaches $A=15$, a large spherical bubble appears for a long lifetime. As the amplitude increases, a cylindrical bubble or phase separation occurs. Large-scale simulations revealed that bubbles emerge before phase separation occurs.}%
\end{figure}

Figure~\ref{fig:13} shows the density field. The simulation box is divided into cells with the dimensions $l_x\times l_y\times l_z = 5\times 5\times 5$ (Movies 5 and 6 in the supplementary material, corresponding to Figs.~\ref{fig:13}(b) and (c), respectively). At $A=5$, the entire system is in a liquid state without gas. For $A=10$, the gas state of the cell occasionally appears in some places. These bubble nuclei do not grow and quickly disappear. As the amplitude further increases, at $A=15$, an obvious bubble emerges close to the oscillating wall. When the amplitude reaches $A=20$, a cylindrical bubble appears, and when $A=40$, a gas layer parallel to the wall is formed, as seen in the case of a small system. Therefore, increasing the size of the system allowed us to observe the dynamics of spherical bubbles, which could not be captured in the small system.

We investigated whether the soundwaves can be described by Burgers' equation when bubbles exist in the system. Figure~\ref{fig:12} shows the results of the numerical solutions of Burgers' equation. Although Burgers' equation and MD results agree well for small amplitudes $A \le 10$, the results exhibit a poor fit for larger amplitudes, where bubbles or gas phases appear. Therefore, Burgers' equation fails to describe a system with phase separation.

\begin{figure}
    \includegraphics[width=3.37in]{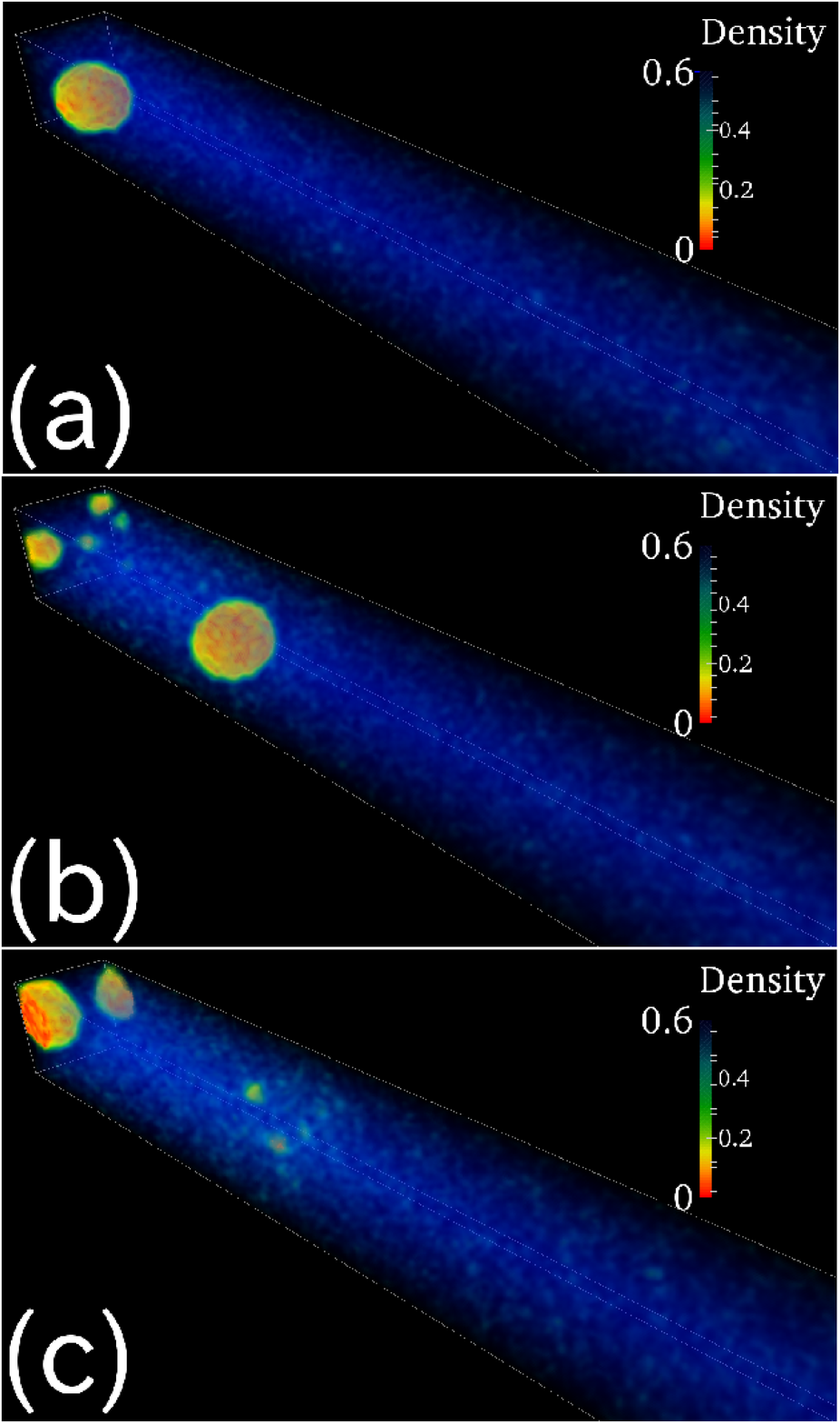}%
    \caption{\label{fig:14} Bubble dynamics from the inception to the collapse in the first-order transition point ($\rho_{0}=0.6$, $T_{0}=0.85$) at the oscillation amplitude $A=15$. (a) A bubble generation near the oscillating wall, (b) bubble traveling toward a wave propagation direction and bubble nucleation, (c) bubble collapse, and new bubble generation and growth. The bubbles repeat a similar cycle from generation to collapse.
    }%
\end{figure}

Figure~\ref{fig:14} and Movie 6 (supplementary material) show the time evolution of the bubbles in the system with $A=15$. A bubble appears owing to the generation and growth of the bubble nuclei near the wall. When the bubble grows sufficiently, it moves in the direction of the wave propagation, and begins to deflate after traveling for a certain distance. Simultaneously, new bubbles appear near the wall. The distal bubble eventually disappears, and the subsequent bubble near the wall grows, moves, and disappears. Thus, the bubble repeats the cycles of formation, growth, movement, and disappearance.

\begin{figure}
    \includegraphics[width=3.37in]{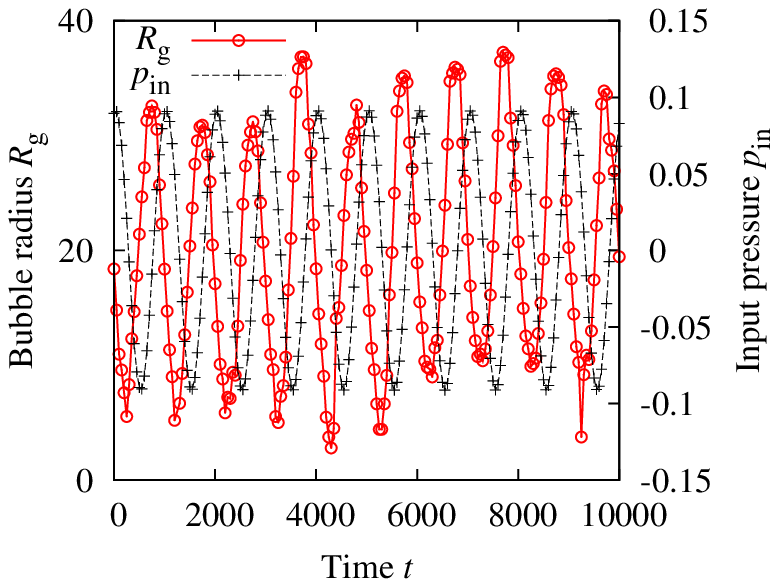}%
    \caption{\label{fig:15} Time evolution of the gyration radius $R_{\rm rg}$ of the largest bubble generated in the first-order transition point ($\rho_{0}=0.6$, $T_{0}=0.85$) at the oscillation amplitude $A=15$.  The amplitude of the pressure wave $p_{\rm in}$ that reached the bubble position is also shown. The oscillations of $R_{\rm g}$ and $p_{\rm in}$ are in antiphase.}%
\end{figure}

To study the time evolution of the bubble size, we estimate the gyration radius $R_{\rm g}$ as follows:
\begin{eqnarray}
    R_{\rm g}^2 &=& \frac{1}{n_{\rm bc}}\sum_{i}({\bm x}_{i}^{\rm b}-{\bm x}^{{\rm b}})^2 ,\label{eq:rg}\\
    {\bm x}^{\rm b} &=& \frac{1}{n_{\rm bc}} \sum_{i}{\bm x}_{i}^{{\rm b}},\label{eq:xb}
\end{eqnarray}
where $n_{\rm bc}$ is the number of cells in a bubble, $x^{\rm b}$ is the geometric center position of the bubble, and ${\bm x}_{i}^{\rm b}$ is the center position of the $i$th cell of the bubble. The summations of Eqs.~(\ref{eq:rg}) and (\ref{eq:xb}) are taken for the cells contained in a given bubble. A cell with a density less than or equal to $(\rho_{\rm l}+\rho_{\rm g})/2=0.34$ is defined to be in the gaseous state. $\rho_{\rm l}$ and $\rho_{\rm g}$ are the liquid and gas phase densities at the gas--liquid coexisting state at $T_{0}=0.85$, respectively. Cells are defined to be bound if a gaseous cell is adjacent to another gaseous cell. Moreover, a bubble is defined to be a cluster of bonded bubble cells. Figure~\ref{fig:15} shows the time evolution of the maximum $R_{\rm g}$ (i.e., $R_{\rm g}$ of the largest bubble) in the system. The pressure wave emitted from the oscillating wall at a time $\Delta \tau = x^{\rm b}/c_{S}$ ago reaches the position of the bubble. Figure~\ref{fig:15} shows the amplitude $p_{\rm in}$ of the pressure wave at its emission.  The bubble repeatedly expands and contracts owing to the pressure waves, such that  $R_{\rm g}$ and $p_{\rm in}$ exhibit oscillations that are synchronized in the antiphase.

\begin{figure*}
    \includegraphics[width=6.69in]{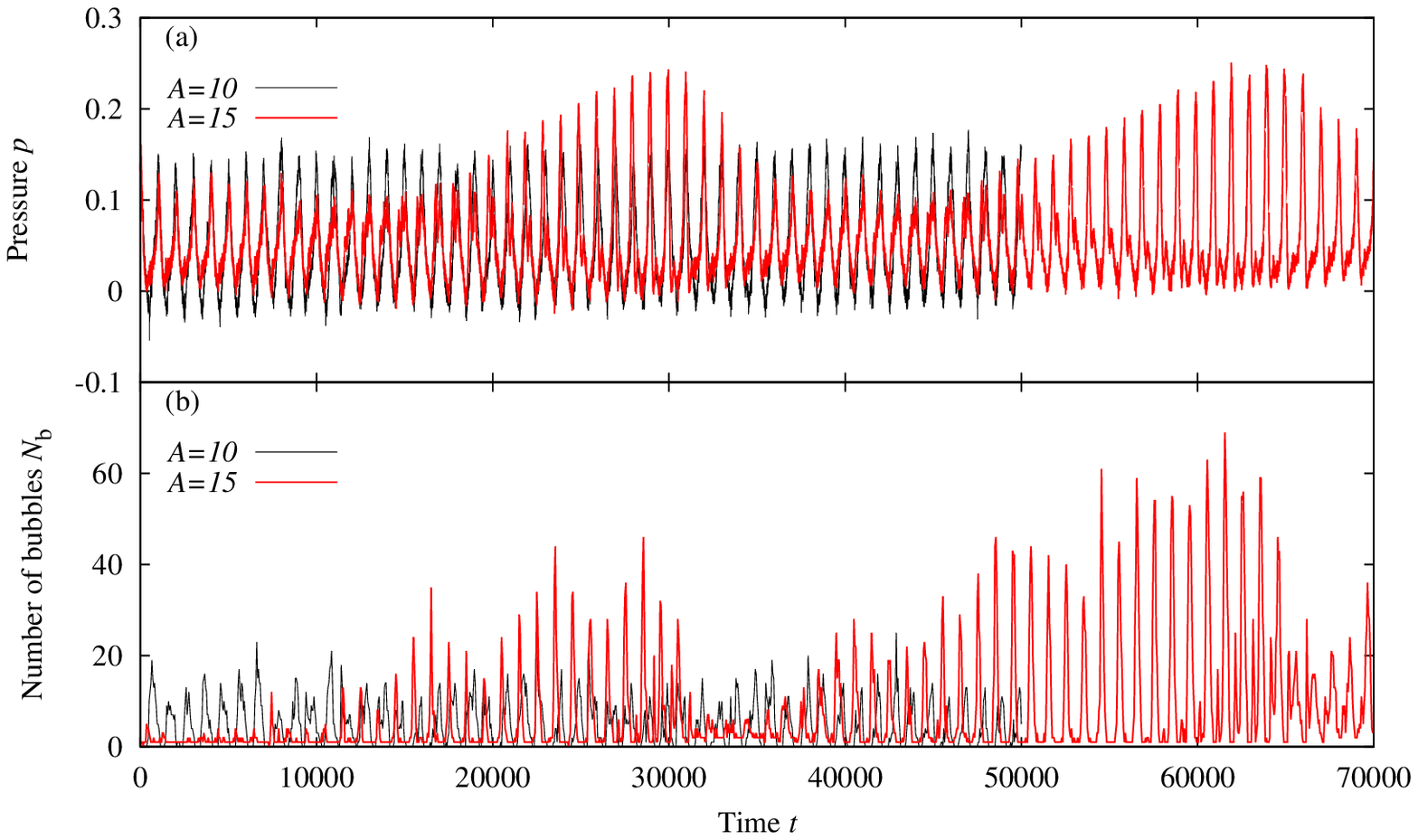}%
    \caption{\label{fig:16} Time evolution of (a) the pressure $p$ near the oscillating wall and (b) the total number of bubbles $N_{\rm b}$ in the first-order transition point ($\rho_{0}=0.6$, $T_{0}=0.85$). The amplitudes of the oscillating wall are $A=10$ and $15$. The pressure is estimated by applying the virial theorem to the slice near the wall, where the slice volume $V_{\rm slice} = 25\times 100 \times 100=250~000$. For $A=10$, the pressure oscillates at a constant amplitude. Moreover, the frequency of bubble generation is almost unchanged. In contrast, for $A=15$, an increase and decrease in the pressure wave amplitude occur. As the amplitude increases, the frequency of bubble generation also increases. The dynamics of nucleation and growth during bubble generation affect the pressure wave.}%
\end{figure*}

As shown in Fig.~\ref{fig:14}, the period from bubble generation to disappearance is relatively long. We investigate the effect of the bubble motion on the pressure waves generated by the oscillating wall in this period. Figure~\ref{fig:16}(a) shows the time evolution of the pressure in the range of $25 \le x < 50$ over approximately two cycles of bubbles. We calculate the pressure using the virial theorem. At $A=10$, whereby small bubbles are generated, the pressure oscillates with a relatively constant amplitude. However, when large bubbles appear ($A=15$), the amplitude increases and decreases following a slow cycle up to $\sim 30~000$. Figure~\ref{fig:16}(b) shows the time evolution of the total number of bubbles in the system. At $A=10$, bubbles repeatedly appear and disappear, and their amplitudes fluctuate randomly. Meanwhile, with $A=15$, the number of bubbles rapidly increases at $t\simeq 10~000$ and $40~000$. This increase is attributed to the bubble formation near the oscillating wall accompanied by an increase in the pressure amplitude, as shown in Fig.~\ref{fig:16}(a). Therefore, the dynamics during bubble generation have a significant effect on the pressure wave.

\section{Summary and Discussion}
In this study, we investigated the effects of gas--liquid phase transitions on soundwaves. The phenomena observed in the first-order and continuous transition regions are significantly different. For the first-order transition region, the amplitude of the soundwave rapidly decays under slight changes in the temperature or amplitude of the oscillating wall. This is attributed to the first-order transition involving a discontinuous change in the density field. The rapid decay of the soundwave at the gas--liquid interface reflects the high attenuation in the gas phase. Meanwhile, for the continuous transition region, the changes are only continuous even with the temperature sensitivity of the soundwaves. Moreover, unlike the first-order transition region, phase separation does not clearly occur. Furthermore, the amplitude of the density fluctuations rapidly increases near the boiling point. Although the fluctuation amplification is attributed to the critical anomaly, it does not affect the waveform. Therefore, the soundwaves depend only on the thermodynamic state in the continuous transition region.

To quantitatively understand the effects of the gas--liquid phase transitions on the acoustic properties, we fitted the results of the MD simulation and Burgers' equation to obtain the attenuation and nonlinear parameters. The numerical solution is compatible with the MD results, except for phase-separated systems. Meanwhile, the attenuation parameters exhibit a divergence tendency in the continuous transition region, reflecting a critical anomaly. In the first-order transition region, a small amount of bubbles are generated before the steady phase separation. Although the effect is not remarkable for critical anomalies, the precursors of phase transitions are expected to affect the acoustic properties. Thus, we demonstrated the estimation of the acoustic properties from the waveform obtained by MD simulation. 

We performed large-scale simulations to examine the effects of the bubbles on the soundwaves. A bubble appears in the vicinity of the oscillating wall, grows, and moves in the sound propagation direction with its radius oscillating. The bubble disappears at a specific distance from the wall, while a new bubble appears near the wall.  A detailed analysis of the bubble motion reveals the interaction between the soundwaves and bubbles (first Bjerknes force~\cite{crum75,louisnard08}). Furthermore, we found that the dynamics of the bubble generation affect the pressure wave. The growth dynamics at the initial stage of the bubble generation of several bubble nuclei in the vicinity of the wall affect the pressure wave amplitude. For the first time, the Bjerknes forces are observed on a molecular scale. Further investigation of the bubble dynamics will help us to better understand ultrasonic cavitation.

In this study, there was no effect of bubble generation on the wavelength, and consequently, the speed of sound did not change. According to the oscillating behavior of the bubbles, the frequency of the oscillating wall is lower than the natural frequency of the bubbles. Therefore, the speed of sound should decrease owing to the appearance of bubbles~\cite{silberman57, wrc05}. However, unlike the experimental results, only a single bubble is observed in this study. Therefore, a large number of bubbles in the system is needed to observe the change in speed of sound. Hence, a larger simulation and analysis should clarify these effects, which would also elucidate on critical issues related to acoustic cavitation, such as interaction between bubbles.

The MD simulation in this study allowed us to observe the phase transition, bubble growth, and bubble--sound interaction without using any phenomenological models. The analysis of the interaction between the bubbles and soundwaves based on the molecular theory is a powerful tool for the justification of hydrodynamic models. Therefore, our approach has a significant potential for a deeper understanding of flows and soundwaves involving phase transitions.

\section*{SUPPLEMENTARY MATERIAL}
See the supplementary material for the movies of the time evolution of the density field.
 
\begin{acknowledgments}
    This research was supported by  JSPS KAKENHI, Grant No. JP15K05201 and JP19H05718. We also acknowledge the Supercomputer Center, Institute for Solid State Physics (ISSP), University of Tokyo, and the Research Center for Computational Science (RCCS), Okazaki, Japan, for allowing us to use their supercomputers.
\end{acknowledgments}

\section*{Data Availability Statement}
The data that support the findings of this study are available from the corresponding author upon reasonable request.


\nocite{*}

\begin{thebibliography}{57}%
\makeatletter
\providecommand \@ifxundefined [1]{%
 \@ifx{#1\undefined}
}%
\providecommand \@ifnum [1]{%
 \ifnum #1\expandafter \@firstoftwo
 \else \expandafter \@secondoftwo
 \fi
}%
\providecommand \@ifx [1]{%
 \ifx #1\expandafter \@firstoftwo
 \else \expandafter \@secondoftwo
 \fi
}%
\providecommand \natexlab [1]{#1}%
\providecommand \enquote  [1]{``#1''}%
\providecommand \bibnamefont  [1]{#1}%
\providecommand \bibfnamefont [1]{#1}%
\providecommand \citenamefont [1]{#1}%
\providecommand \href@noop [0]{\@secondoftwo}%
\providecommand \href [0]{\begingroup \@sanitize@url \@href}%
\providecommand \@href[1]{\@@startlink{#1}\@@href}%
\providecommand \@@href[1]{\endgroup#1\@@endlink}%
\providecommand \@sanitize@url [0]{\catcode `\\12\catcode `\$12\catcode
  `\&12\catcode `\#12\catcode `\^12\catcode `\_12\catcode `\%12\relax}%
\providecommand \@@startlink[1]{}%
\providecommand \@@endlink[0]{}%
\providecommand \url  [0]{\begingroup\@sanitize@url \@url }%
\providecommand \@url [1]{\endgroup\@href {#1}{\urlprefix }}%
\providecommand \urlprefix  [0]{URL }%
\providecommand \Eprint [0]{\href }%
\providecommand \doibase [0]{http://dx.doi.org/}%
\providecommand \selectlanguage [0]{\@gobble}%
\providecommand \bibinfo  [0]{\@secondoftwo}%
\providecommand \bibfield  [0]{\@secondoftwo}%
\providecommand \translation [1]{[#1]}%
\providecommand \BibitemOpen [0]{}%
\providecommand \bibitemStop [0]{}%
\providecommand \bibitemNoStop [0]{.\EOS\space}%
\providecommand \EOS [0]{\spacefactor3000\relax}%
\providecommand \BibitemShut  [1]{\csname bibitem#1\endcsname}%
\let\auto@bib@innerbib\@empty
\bibitem [{\citenamefont {Thompson}\ and\ \citenamefont
  {Doraiswamy}(1999)}]{td99}%
  \BibitemOpen
  \bibfield  {author} {\bibinfo {author} {\bibfnamefont {L.~H.}\ \bibnamefont
  {Thompson}}\ and\ \bibinfo {author} {\bibfnamefont {L.}~\bibnamefont
  {Doraiswamy}},\ }\bibfield  {title} {\enquote {\bibinfo {title}
  {Sonochemistry: science and engineering},}\ }\href@noop {} {\bibfield
  {journal} {\bibinfo  {journal} {Ind. Eng. Chem. Res.}\ }\textbf {\bibinfo
  {volume} {38}},\ \bibinfo {pages} {1215--1249} (\bibinfo {year}
  {1999})}\BibitemShut {NoStop}%
\bibitem [{\citenamefont {Lauterborn}\ \emph {et~al.}(2007)\citenamefont
  {Lauterborn}, \citenamefont {Kurz}, \citenamefont {Geisler}, \citenamefont
  {Schanz},\ and\ \citenamefont {Lindau}}]{lkg07}%
  \BibitemOpen
  \bibfield  {author} {\bibinfo {author} {\bibfnamefont {W.}~\bibnamefont
  {Lauterborn}}, \bibinfo {author} {\bibfnamefont {T.}~\bibnamefont {Kurz}},
  \bibinfo {author} {\bibfnamefont {R.}~\bibnamefont {Geisler}}, \bibinfo
  {author} {\bibfnamefont {D.}~\bibnamefont {Schanz}}, \ and\ \bibinfo {author}
  {\bibfnamefont {O.}~\bibnamefont {Lindau}},\ }\bibfield  {title} {\enquote
  {\bibinfo {title} {Acoustic cavitation, bubble dynamics and
  sonoluminescence},}\ }\href@noop {} {\bibfield  {journal} {\bibinfo
  {journal} {Ultrason. Sonochem.}\ }\textbf {\bibinfo {volume} {14}},\ \bibinfo
  {pages} {484--491} (\bibinfo {year} {2007})}\BibitemShut {NoStop}%
\bibitem [{\citenamefont {Kentish}\ and\ \citenamefont {Feng}(2014)}]{kf14}%
  \BibitemOpen
  \bibfield  {author} {\bibinfo {author} {\bibfnamefont {S.}~\bibnamefont
  {Kentish}}\ and\ \bibinfo {author} {\bibfnamefont {H.}~\bibnamefont {Feng}},\
  }\bibfield  {title} {\enquote {\bibinfo {title} {Applications of power
  ultrasound in food processing},}\ }\href@noop {} {\bibfield  {journal}
  {\bibinfo  {journal} {Annu. Rev. Food Sci.}\ }\textbf {\bibinfo {volume}
  {5}},\ \bibinfo {pages} {263--284} (\bibinfo {year} {2014})}\BibitemShut
  {NoStop}%
\bibitem [{\citenamefont {Wood}\ and\ \citenamefont {Sehgal}(2015)}]{ws15}%
  \BibitemOpen
  \bibfield  {author} {\bibinfo {author} {\bibfnamefont {A.~K.~W.}\
  \bibnamefont {Wood}}\ and\ \bibinfo {author} {\bibfnamefont {C.~M.}\
  \bibnamefont {Sehgal}},\ }\bibfield  {title} {\enquote {\bibinfo {title} {A
  review of low-intensity ultrasound for cancer therapy},}\ }\href@noop {}
  {\bibfield  {journal} {\bibinfo  {journal} {Ultrasound Med. Biol.}\ }\textbf
  {\bibinfo {volume} {41}},\ \bibinfo {pages} {905--928} (\bibinfo {year}
  {2015})}\BibitemShut {NoStop}%
\bibitem [{\citenamefont {Mason}(2016)}]{mason16}%
  \BibitemOpen
  \bibfield  {author} {\bibinfo {author} {\bibfnamefont {T.~J.}\ \bibnamefont
  {Mason}},\ }\bibfield  {title} {\enquote {\bibinfo {title} {Ultrasonic
  cleaning: An historical perspective},}\ }\href@noop {} {\bibfield  {journal}
  {\bibinfo  {journal} {Ultrason. Sonochem.}\ }\textbf {\bibinfo {volume}
  {29}},\ \bibinfo {pages} {519--523} (\bibinfo {year} {2016})}\BibitemShut
  {NoStop}%
\bibitem [{\citenamefont {Eskin}\ \emph {et~al.}(2019)\citenamefont {Eskin},
  \citenamefont {Tzanakis}, \citenamefont {Wang}, \citenamefont {Lebon},
  \citenamefont {Subroto}, \citenamefont {Pericleous},\ and\ \citenamefont
  {Mi}}]{etw19}%
  \BibitemOpen
  \bibfield  {author} {\bibinfo {author} {\bibfnamefont {D.~G.}\ \bibnamefont
  {Eskin}}, \bibinfo {author} {\bibfnamefont {I.}~\bibnamefont {Tzanakis}},
  \bibinfo {author} {\bibfnamefont {F.}~\bibnamefont {Wang}}, \bibinfo {author}
  {\bibfnamefont {G.~S.~B.}\ \bibnamefont {Lebon}}, \bibinfo {author}
  {\bibfnamefont {T.}~\bibnamefont {Subroto}}, \bibinfo {author} {\bibfnamefont
  {K.}~\bibnamefont {Pericleous}}, \ and\ \bibinfo {author} {\bibfnamefont
  {J.}~\bibnamefont {Mi}},\ }\bibfield  {title} {\enquote {\bibinfo {title}
  {Fundamental studies of ultrasonic melt processing},}\ }\href@noop {}
  {\bibfield  {journal} {\bibinfo  {journal} {Ultrason. Sonochem.}\ }\textbf
  {\bibinfo {volume} {52}},\ \bibinfo {pages} {455--467} (\bibinfo {year}
  {2019})}\BibitemShut {NoStop}%
\bibitem [{\citenamefont {Izadifar}, \citenamefont {Babyn},\ and\ \citenamefont
  {Chapman}(2019)}]{ibc19}%
  \BibitemOpen
  \bibfield  {author} {\bibinfo {author} {\bibfnamefont {Z.}~\bibnamefont
  {Izadifar}}, \bibinfo {author} {\bibfnamefont {P.}~\bibnamefont {Babyn}}, \
  and\ \bibinfo {author} {\bibfnamefont {D.}~\bibnamefont {Chapman}},\
  }\bibfield  {title} {\enquote {\bibinfo {title} {Ultrasound
  cavitation/microbubble detection and medical applications},}\ }\href@noop {}
  {\bibfield  {journal} {\bibinfo  {journal} {J. Med. Biol. Eng.}\ }\textbf
  {\bibinfo {volume} {39}},\ \bibinfo {pages} {259--276} (\bibinfo {year}
  {2019})}\BibitemShut {NoStop}%
\bibitem [{\citenamefont {Al-Jawadi}\ and\ \citenamefont
  {Thakur}(2020)}]{at20}%
  \BibitemOpen
  \bibfield  {author} {\bibinfo {author} {\bibfnamefont {S.}~\bibnamefont
  {Al-Jawadi}}\ and\ \bibinfo {author} {\bibfnamefont {S.~S.}\ \bibnamefont
  {Thakur}},\ }\bibfield  {title} {\enquote {\bibinfo {title}
  {Ultrasound-responsive lipid microbubbles for drug delivery: A review of
  preparation techniques to optimise formulation size, stability and drug
  loading},}\ }\href@noop {} {\bibfield  {journal} {\bibinfo  {journal} {Int.
  J. Pharm.}\ }\textbf {\bibinfo {volume} {585}},\ \bibinfo {pages} {119559}
  (\bibinfo {year} {2020})}\BibitemShut {NoStop}%
\bibitem [{\citenamefont {Savun-Hekimo{\u{g}}lu}(2020)}]{savun20}%
  \BibitemOpen
  \bibfield  {author} {\bibinfo {author} {\bibfnamefont {B.}~\bibnamefont
  {Savun-Hekimo{\u{g}}lu}},\ }\bibfield  {title} {\enquote {\bibinfo {title} {A
  review on sonochemistry and its environmental applications},}\ }in\
  \href@noop {} {\emph {\bibinfo {booktitle} {Acoustics}}},\ Vol.~\bibinfo
  {volume} {2}\ (\bibinfo {organization} {Multidisciplinary Digital Publishing
  Institute},\ \bibinfo {year} {2020})\ pp.\ \bibinfo {pages}
  {766--775}\BibitemShut {NoStop}%
\bibitem [{\citenamefont {Soyama}(2020)}]{soyama20}%
  \BibitemOpen
  \bibfield  {author} {\bibinfo {author} {\bibfnamefont {H.}~\bibnamefont
  {Soyama}},\ }\bibfield  {title} {\enquote {\bibinfo {title} {Cavitation
  peening: A review},}\ }\href@noop {} {\bibfield  {journal} {\bibinfo
  {journal} {Metals}\ }\textbf {\bibinfo {volume} {10}},\ \bibinfo {pages}
  {270} (\bibinfo {year} {2020})}\BibitemShut {NoStop}%
\bibitem [{\citenamefont {Yao}, \citenamefont {Pan},\ and\ \citenamefont
  {Liu}(2020)}]{ypl20}%
  \BibitemOpen
  \bibfield  {author} {\bibinfo {author} {\bibfnamefont {Y.}~\bibnamefont
  {Yao}}, \bibinfo {author} {\bibfnamefont {Y.}~\bibnamefont {Pan}}, \ and\
  \bibinfo {author} {\bibfnamefont {S.}~\bibnamefont {Liu}},\ }\bibfield
  {title} {\enquote {\bibinfo {title} {Power ultrasound and its applications: A
  state-of-the-art review},}\ }\href@noop {} {\bibfield  {journal} {\bibinfo
  {journal} {Ultrason. Sonochem.}\ }\textbf {\bibinfo {volume} {62}},\ \bibinfo
  {pages} {104722} (\bibinfo {year} {2020})}\BibitemShut {NoStop}%
\bibitem [{\citenamefont {Lauterborn}\ and\ \citenamefont {Ohl}(1997)}]{wc97}%
  \BibitemOpen
  \bibfield  {author} {\bibinfo {author} {\bibfnamefont {W.}~\bibnamefont
  {Lauterborn}}\ and\ \bibinfo {author} {\bibfnamefont {C.-D.}\ \bibnamefont
  {Ohl}},\ }\bibfield  {title} {\enquote {\bibinfo {title} {Cavitation bubble
  dynamics},}\ }\href
  {https://www.sciencedirect.com/science/article/pii/S1350417797000096}
  {\bibfield  {journal} {\bibinfo  {journal} {Ultrason. Sonochem.}\ }\textbf
  {\bibinfo {volume} {4}},\ \bibinfo {pages} {65--75} (\bibinfo {year}
  {1997})}\BibitemShut {NoStop}%
\bibitem [{\citenamefont {Moussatov}, \citenamefont {Granger},\ and\
  \citenamefont {Dubus}(2003)}]{mgd03}%
  \BibitemOpen
  \bibfield  {author} {\bibinfo {author} {\bibfnamefont {A.}~\bibnamefont
  {Moussatov}}, \bibinfo {author} {\bibfnamefont {C.}~\bibnamefont {Granger}},
  \ and\ \bibinfo {author} {\bibfnamefont {B.}~\bibnamefont {Dubus}},\
  }\bibfield  {title} {\enquote {\bibinfo {title} {Cone-like bubble formation
  in ultrasonic cavitation field},}\ }\href@noop {} {\bibfield  {journal}
  {\bibinfo  {journal} {Ultrason. Sonochem.}\ }\textbf {\bibinfo {volume}
  {10}},\ \bibinfo {pages} {191--195} (\bibinfo {year} {2003})}\BibitemShut
  {NoStop}%
\bibitem [{\citenamefont {Dubus}\ \emph {et~al.}(2010)\citenamefont {Dubus},
  \citenamefont {Vanhille}, \citenamefont {Campos-Pozuelo},\ and\ \citenamefont
  {Granger}}]{dvc10}%
  \BibitemOpen
  \bibfield  {author} {\bibinfo {author} {\bibfnamefont {B.}~\bibnamefont
  {Dubus}}, \bibinfo {author} {\bibfnamefont {C.}~\bibnamefont {Vanhille}},
  \bibinfo {author} {\bibfnamefont {C.}~\bibnamefont {Campos-Pozuelo}}, \ and\
  \bibinfo {author} {\bibfnamefont {C.}~\bibnamefont {Granger}},\ }\bibfield
  {title} {\enquote {\bibinfo {title} {On the physical origin of conical bubble
  structure under an ultrasonic horn},}\ }\href@noop {} {\bibfield  {journal}
  {\bibinfo  {journal} {Ultrason. Sonochem.}\ }\textbf {\bibinfo {volume}
  {17}},\ \bibinfo {pages} {810--818} (\bibinfo {year} {2010})}\BibitemShut
  {NoStop}%
\bibitem [{\citenamefont {{\v{Z}}nidar{\v{c}}i{\v{c}}}\ \emph
  {et~al.}(2014)\citenamefont {{\v{Z}}nidar{\v{c}}i{\v{c}}}, \citenamefont
  {Mettin}, \citenamefont {Cair{\'o}s},\ and\ \citenamefont {Dular}}]{zmc14}%
  \BibitemOpen
  \bibfield  {author} {\bibinfo {author} {\bibfnamefont {A.}~\bibnamefont
  {{\v{Z}}nidar{\v{c}}i{\v{c}}}}, \bibinfo {author} {\bibfnamefont
  {R.}~\bibnamefont {Mettin}}, \bibinfo {author} {\bibfnamefont
  {C.}~\bibnamefont {Cair{\'o}s}}, \ and\ \bibinfo {author} {\bibfnamefont
  {M.}~\bibnamefont {Dular}},\ }\bibfield  {title} {\enquote {\bibinfo {title}
  {Attached cavitation at a small diameter ultrasonic horn tip},}\ }\href@noop
  {} {\bibfield  {journal} {\bibinfo  {journal} {Phys. Fluids}\ }\textbf
  {\bibinfo {volume} {26}},\ \bibinfo {pages} {023304} (\bibinfo {year}
  {2014})}\BibitemShut {NoStop}%
\bibitem [{\citenamefont {Tzanakis}\ \emph {et~al.}(2017)\citenamefont
  {Tzanakis}, \citenamefont {Lebon}, \citenamefont {Eskin},\ and\ \citenamefont
  {Pericleous}}]{tle17}%
  \BibitemOpen
  \bibfield  {author} {\bibinfo {author} {\bibfnamefont {I.}~\bibnamefont
  {Tzanakis}}, \bibinfo {author} {\bibfnamefont {G.}~\bibnamefont {Lebon}},
  \bibinfo {author} {\bibfnamefont {D.}~\bibnamefont {Eskin}}, \ and\ \bibinfo
  {author} {\bibfnamefont {K.}~\bibnamefont {Pericleous}},\ }\bibfield  {title}
  {\enquote {\bibinfo {title} {Characterizing the cavitation development and
  acoustic spectrum in various liquids},}\ }\href@noop {} {\bibfield  {journal}
  {\bibinfo  {journal} {Ultrason. Sonochem.}\ }\textbf {\bibinfo {volume}
  {34}},\ \bibinfo {pages} {651--662} (\bibinfo {year} {2017})}\BibitemShut
  {NoStop}%
\bibitem [{\citenamefont {Fang}, \citenamefont {Yamamoto},\ and\ \citenamefont
  {Komarov}(2018)}]{fyk18}%
  \BibitemOpen
  \bibfield  {author} {\bibinfo {author} {\bibfnamefont {Y.}~\bibnamefont
  {Fang}}, \bibinfo {author} {\bibfnamefont {T.}~\bibnamefont {Yamamoto}}, \
  and\ \bibinfo {author} {\bibfnamefont {S.}~\bibnamefont {Komarov}},\
  }\bibfield  {title} {\enquote {\bibinfo {title} {Cavitation and acoustic
  streaming generated by different sonotrode tips},}\ }\href@noop {} {\bibfield
   {journal} {\bibinfo  {journal} {Ultrason. Sonochem.}\ }\textbf {\bibinfo
  {volume} {48}},\ \bibinfo {pages} {79--87} (\bibinfo {year}
  {2018})}\BibitemShut {NoStop}%
\bibitem [{\citenamefont {Yusuf}, \citenamefont {Symes},\ and\ \citenamefont
  {Prentice}(2021)}]{ysp21}%
  \BibitemOpen
  \bibfield  {author} {\bibinfo {author} {\bibfnamefont {L.}~\bibnamefont
  {Yusuf}}, \bibinfo {author} {\bibfnamefont {M.~D.}\ \bibnamefont {Symes}}, \
  and\ \bibinfo {author} {\bibfnamefont {P.}~\bibnamefont {Prentice}},\
  }\bibfield  {title} {\enquote {\bibinfo {title} {Characterising the
  cavitation activity generated by an ultrasonic horn at varying tip-vibration
  amplitudes},}\ }\href@noop {} {\bibfield  {journal} {\bibinfo  {journal}
  {Ultrason. Sonochem.}\ }\textbf {\bibinfo {volume} {70}},\ \bibinfo {pages}
  {105273} (\bibinfo {year} {2021})}\BibitemShut {NoStop}%
\bibitem [{\citenamefont {Hodnett}, \citenamefont {Chow},\ and\ \citenamefont
  {Zeqiri}(2004)}]{hcz04}%
  \BibitemOpen
  \bibfield  {author} {\bibinfo {author} {\bibfnamefont {M.}~\bibnamefont
  {Hodnett}}, \bibinfo {author} {\bibfnamefont {R.}~\bibnamefont {Chow}}, \
  and\ \bibinfo {author} {\bibfnamefont {B.}~\bibnamefont {Zeqiri}},\
  }\bibfield  {title} {\enquote {\bibinfo {title} {High-frequency acoustic
  emissions generated by a 20 khz sonochemical horn processor detected using a
  novel broadband acoustic sensor: a preliminary study},}\ }\href@noop {}
  {\bibfield  {journal} {\bibinfo  {journal} {Ultrason. Sonochem.}\ }\textbf
  {\bibinfo {volume} {11}},\ \bibinfo {pages} {441--454} (\bibinfo {year}
  {2004})}\BibitemShut {NoStop}%
\bibitem [{\citenamefont {Tan}\ and\ \citenamefont {Yeo}(2019)}]{ty19}%
  \BibitemOpen
  \bibfield  {author} {\bibinfo {author} {\bibfnamefont {K.}~\bibnamefont
  {Tan}}\ and\ \bibinfo {author} {\bibfnamefont {S.}~\bibnamefont {Yeo}},\
  }\bibfield  {title} {\enquote {\bibinfo {title} {Bubble dynamics and
  cavitation intensity in milli-scale channels under an ultrasonic horn},}\
  }\href@noop {} {\bibfield  {journal} {\bibinfo  {journal} {Ultrason.
  Sonochem.}\ }\textbf {\bibinfo {volume} {58}},\ \bibinfo {pages} {104666}
  (\bibinfo {year} {2019})}\BibitemShut {NoStop}%
\bibitem [{\citenamefont {Silberman}(1957)}]{silberman57}%
  \BibitemOpen
  \bibfield  {author} {\bibinfo {author} {\bibfnamefont {E.}~\bibnamefont
  {Silberman}},\ }\bibfield  {title} {\enquote {\bibinfo {title} {Sound
  velocity and attenuation in bubbly mixtures measured in standing wave
  tubes},}\ }\href@noop {} {\bibfield  {journal} {\bibinfo  {journal} {J.
  Acoust. Soc. Am.}\ }\textbf {\bibinfo {volume} {29}},\ \bibinfo {pages}
  {925--933} (\bibinfo {year} {1957})}\BibitemShut {NoStop}%
\bibitem [{\citenamefont {Wilson}, \citenamefont {Roy},\ and\ \citenamefont
  {Carey}(2005)}]{wrc05}%
  \BibitemOpen
  \bibfield  {author} {\bibinfo {author} {\bibfnamefont {P.~S.}\ \bibnamefont
  {Wilson}}, \bibinfo {author} {\bibfnamefont {R.~A.}\ \bibnamefont {Roy}}, \
  and\ \bibinfo {author} {\bibfnamefont {W.~M.}\ \bibnamefont {Carey}},\
  }\bibfield  {title} {\enquote {\bibinfo {title} {Phase speed and attenuation
  in bubbly liquids inferred from impedance measurements near the individual
  bubble resonance frequency},}\ }\href@noop {} {\bibfield  {journal} {\bibinfo
   {journal} {J. Acoust. Soc. Am.}\ }\textbf {\bibinfo {volume} {117}},\
  \bibinfo {pages} {1895--1910} (\bibinfo {year} {2005})}\BibitemShut {NoStop}%
\bibitem [{\citenamefont {Kanagawa}\ \emph {et~al.}(2010)\citenamefont
  {Kanagawa}, \citenamefont {Yano}, \citenamefont {Watanabe},\ and\
  \citenamefont {Fujikawa}}]{kyw10}%
  \BibitemOpen
  \bibfield  {author} {\bibinfo {author} {\bibfnamefont {T.}~\bibnamefont
  {Kanagawa}}, \bibinfo {author} {\bibfnamefont {T.}~\bibnamefont {Yano}},
  \bibinfo {author} {\bibfnamefont {M.}~\bibnamefont {Watanabe}}, \ and\
  \bibinfo {author} {\bibfnamefont {S.}~\bibnamefont {Fujikawa}},\ }\bibfield
  {title} {\enquote {\bibinfo {title} {Unified theory based on parameter
  scaling for derivation of nonlinear wave equations in bubbly liquids},}\
  }\href@noop {} {\bibfield  {journal} {\bibinfo  {journal} {J. Fluid Sci.
  Technol.}\ }\textbf {\bibinfo {volume} {5}},\ \bibinfo {pages} {351--369}
  (\bibinfo {year} {2010})}\BibitemShut {NoStop}%
\bibitem [{\citenamefont {Louisnard}(2012)}]{louisnard12}%
  \BibitemOpen
  \bibfield  {author} {\bibinfo {author} {\bibfnamefont {O.}~\bibnamefont
  {Louisnard}},\ }\bibfield  {title} {\enquote {\bibinfo {title} {A simple
  model of ultrasound propagation in a cavitating liquid. part i: Theory,
  nonlinear attenuation and traveling wave generation},}\ }\href@noop {}
  {\bibfield  {journal} {\bibinfo  {journal} {Ultrason. Sonochem.}\ }\textbf
  {\bibinfo {volume} {19}},\ \bibinfo {pages} {56--65} (\bibinfo {year}
  {2012})}\BibitemShut {NoStop}%
\bibitem [{\citenamefont {Fuster}, \citenamefont {Conoir},\ and\ \citenamefont
  {Colonius}(2014)}]{fcc14}%
  \BibitemOpen
  \bibfield  {author} {\bibinfo {author} {\bibfnamefont {D.}~\bibnamefont
  {Fuster}}, \bibinfo {author} {\bibfnamefont {J.-M.}\ \bibnamefont {Conoir}},
  \ and\ \bibinfo {author} {\bibfnamefont {T.}~\bibnamefont {Colonius}},\
  }\bibfield  {title} {\enquote {\bibinfo {title} {Effect of direct
  bubble-bubble interactions on linear-wave propagation in bubbly liquids},}\
  }\href@noop {} {\bibfield  {journal} {\bibinfo  {journal} {Phys. Rev. E}\
  }\textbf {\bibinfo {volume} {90}},\ \bibinfo {pages} {063010} (\bibinfo
  {year} {2014})}\BibitemShut {NoStop}%
\bibitem [{\citenamefont {Prosperetti}(2015)}]{prosperetti15}%
  \BibitemOpen
  \bibfield  {author} {\bibinfo {author} {\bibfnamefont {A.}~\bibnamefont
  {Prosperetti}},\ }\bibfield  {title} {\enquote {\bibinfo {title} {The speed
  of sound in a gas--vapour bubbly liquid},}\ }\href@noop {} {\bibfield
  {journal} {\bibinfo  {journal} {Interface Focus}\ }\textbf {\bibinfo {volume}
  {5}},\ \bibinfo {pages} {20150024} (\bibinfo {year} {2015})}\BibitemShut
  {NoStop}%
\bibitem [{\citenamefont {Zhang}, \citenamefont {Guo},\ and\ \citenamefont
  {Du}(2018)}]{zgd18}%
  \BibitemOpen
  \bibfield  {author} {\bibinfo {author} {\bibfnamefont {Y.}~\bibnamefont
  {Zhang}}, \bibinfo {author} {\bibfnamefont {Z.}~\bibnamefont {Guo}}, \ and\
  \bibinfo {author} {\bibfnamefont {X.}~\bibnamefont {Du}},\ }\bibfield
  {title} {\enquote {\bibinfo {title} {Wave propagation in liquids with
  oscillating vapor-gas bubbles},}\ }\href@noop {} {\bibfield  {journal}
  {\bibinfo  {journal} {Appl. Therm. Eng.}\ }\textbf {\bibinfo {volume}
  {133}},\ \bibinfo {pages} {483--492} (\bibinfo {year} {2018})}\BibitemShut
  {NoStop}%
\bibitem [{\citenamefont {Zhang}\ \emph {et~al.}(2018)\citenamefont {Zhang},
  \citenamefont {Guo}, \citenamefont {Gao},\ and\ \citenamefont {Du}}]{zgg18}%
  \BibitemOpen
  \bibfield  {author} {\bibinfo {author} {\bibfnamefont {Y.}~\bibnamefont
  {Zhang}}, \bibinfo {author} {\bibfnamefont {Z.}~\bibnamefont {Guo}}, \bibinfo
  {author} {\bibfnamefont {Y.}~\bibnamefont {Gao}}, \ and\ \bibinfo {author}
  {\bibfnamefont {X.}~\bibnamefont {Du}},\ }\bibfield  {title} {\enquote
  {\bibinfo {title} {Acoustic wave propagation in bubbly flow with gas, vapor
  or their mixtures},}\ }\href@noop {} {\bibfield  {journal} {\bibinfo
  {journal} {Ultrason. Sonochem.}\ }\textbf {\bibinfo {volume} {40}},\ \bibinfo
  {pages} {40--45} (\bibinfo {year} {2018})}\BibitemShut {NoStop}%
\bibitem [{\citenamefont {Trujillo}(2018)}]{trujillo18}%
  \BibitemOpen
  \bibfield  {author} {\bibinfo {author} {\bibfnamefont {F.~J.}\ \bibnamefont
  {Trujillo}},\ }\bibfield  {title} {\enquote {\bibinfo {title} {A strict
  formulation of a nonlinear helmholtz equation for the propagation of sound in
  bubbly liquids. part i: Theory and validation at low acoustic pressure
  amplitudes},}\ }\href@noop {} {\bibfield  {journal} {\bibinfo  {journal}
  {Ultrason. Sonochem.}\ }\textbf {\bibinfo {volume} {47}},\ \bibinfo {pages}
  {75--98} (\bibinfo {year} {2018})}\BibitemShut {NoStop}%
\bibitem [{\citenamefont {Trujillo}(2020)}]{trujillo20}%
  \BibitemOpen
  \bibfield  {author} {\bibinfo {author} {\bibfnamefont {F.~J.}\ \bibnamefont
  {Trujillo}},\ }\bibfield  {title} {\enquote {\bibinfo {title} {A strict
  formulation of a nonlinear helmholtz equation for the propagation of sound in
  bubbly liquids. part ii: Application to ultrasonic cavitation},}\ }\href@noop
  {} {\bibfield  {journal} {\bibinfo  {journal} {Ultrason. Sonochem.}\ }\textbf
  {\bibinfo {volume} {65}},\ \bibinfo {pages} {105056} (\bibinfo {year}
  {2020})}\BibitemShut {NoStop}%
\bibitem [{\citenamefont {Crum}(1975)}]{crum75}%
  \BibitemOpen
  \bibfield  {author} {\bibinfo {author} {\bibfnamefont {L.~A.}\ \bibnamefont
  {Crum}},\ }\bibfield  {title} {\enquote {\bibinfo {title} {Bjerknes forces on
  bubbles in a stationary sound field},}\ }\href@noop {} {\bibfield  {journal}
  {\bibinfo  {journal} {J. Acoust. Soc. Am.}\ }\textbf {\bibinfo {volume}
  {57}},\ \bibinfo {pages} {1363--1370} (\bibinfo {year} {1975})}\BibitemShut
  {NoStop}%
\bibitem [{\citenamefont {Louisnard}(2008)}]{louisnard08}%
  \BibitemOpen
  \bibfield  {author} {\bibinfo {author} {\bibfnamefont {O.}~\bibnamefont
  {Louisnard}},\ }\bibfield  {title} {\enquote {\bibinfo {title} {Analytical
  expressions for primary bjerknes force on inertial cavitation bubbles},}\
  }\href@noop {} {\bibfield  {journal} {\bibinfo  {journal} {Phys. Rev. E}\
  }\textbf {\bibinfo {volume} {78}},\ \bibinfo {pages} {036322} (\bibinfo
  {year} {2008})}\BibitemShut {NoStop}%
\bibitem [{\citenamefont {Yasui}\ \emph {et~al.}(2008)\citenamefont {Yasui},
  \citenamefont {Iida}, \citenamefont {Tuziuti}, \citenamefont {Kozuka},\ and\
  \citenamefont {Towata}}]{yit08}%
  \BibitemOpen
  \bibfield  {author} {\bibinfo {author} {\bibfnamefont {K.}~\bibnamefont
  {Yasui}}, \bibinfo {author} {\bibfnamefont {Y.}~\bibnamefont {Iida}},
  \bibinfo {author} {\bibfnamefont {T.}~\bibnamefont {Tuziuti}}, \bibinfo
  {author} {\bibfnamefont {T.}~\bibnamefont {Kozuka}}, \ and\ \bibinfo {author}
  {\bibfnamefont {A.}~\bibnamefont {Towata}},\ }\bibfield  {title} {\enquote
  {\bibinfo {title} {Strongly interacting bubbles under an ultrasonic horn},}\
  }\href@noop {} {\bibfield  {journal} {\bibinfo  {journal} {Phys. Rev. E}\
  }\textbf {\bibinfo {volume} {77}},\ \bibinfo {pages} {016609} (\bibinfo
  {year} {2008})}\BibitemShut {NoStop}%
\bibitem [{\citenamefont {Fuster}\ and\ \citenamefont {Montel}(2015)}]{fm15}%
  \BibitemOpen
  \bibfield  {author} {\bibinfo {author} {\bibfnamefont {D.}~\bibnamefont
  {Fuster}}\ and\ \bibinfo {author} {\bibfnamefont {F.}~\bibnamefont
  {Montel}},\ }\bibfield  {title} {\enquote {\bibinfo {title} {Mass transfer
  effects on linear wave propagation in diluted bubbly liquids},}\ }\href@noop
  {} {\bibfield  {journal} {\bibinfo  {journal} {J. Fluid Mech.}\ }\textbf
  {\bibinfo {volume} {779}},\ \bibinfo {pages} {598--621} (\bibinfo {year}
  {2015})}\BibitemShut {NoStop}%
\bibitem [{\citenamefont {Xu}(2018)}]{xu18}%
  \BibitemOpen
  \bibfield  {author} {\bibinfo {author} {\bibfnamefont {Z.}~\bibnamefont
  {Xu}},\ }\bibfield  {title} {\enquote {\bibinfo {title} {Numerical simulation
  of the coalescence of two bubbles in an ultrasound field},}\ }\href@noop {}
  {\bibfield  {journal} {\bibinfo  {journal} {Ultrason. Sonochem.}\ }\textbf
  {\bibinfo {volume} {49}},\ \bibinfo {pages} {277--282} (\bibinfo {year}
  {2018})}\BibitemShut {NoStop}%
\bibitem [{\citenamefont {Ma}\ \emph {et~al.}(2018)\citenamefont {Ma},
  \citenamefont {Huang}, \citenamefont {Li}, \citenamefont {Chang},
  \citenamefont {Qiu}, \citenamefont {Su}, \citenamefont {Fu},\ and\
  \citenamefont {Wang}}]{mhl18}%
  \BibitemOpen
  \bibfield  {author} {\bibinfo {author} {\bibfnamefont {X.}~\bibnamefont
  {Ma}}, \bibinfo {author} {\bibfnamefont {B.}~\bibnamefont {Huang}}, \bibinfo
  {author} {\bibfnamefont {Y.}~\bibnamefont {Li}}, \bibinfo {author}
  {\bibfnamefont {Q.}~\bibnamefont {Chang}}, \bibinfo {author} {\bibfnamefont
  {S.}~\bibnamefont {Qiu}}, \bibinfo {author} {\bibfnamefont {Z.}~\bibnamefont
  {Su}}, \bibinfo {author} {\bibfnamefont {X.}~\bibnamefont {Fu}}, \ and\
  \bibinfo {author} {\bibfnamefont {G.}~\bibnamefont {Wang}},\ }\bibfield
  {title} {\enquote {\bibinfo {title} {Numerical simulation of single bubble
  dynamics under acoustic travelling waves},}\ }\href@noop {} {\bibfield
  {journal} {\bibinfo  {journal} {Ultrason. Sonochem.}\ }\textbf {\bibinfo
  {volume} {42}},\ \bibinfo {pages} {619--630} (\bibinfo {year}
  {2018})}\BibitemShut {NoStop}%
\bibitem [{\citenamefont {Qiu}\ \emph {et~al.}(2018)\citenamefont {Qiu},
  \citenamefont {Ma}, \citenamefont {Huang}, \citenamefont {Li}, \citenamefont
  {Wang},\ and\ \citenamefont {Zhang}}]{qmh18}%
  \BibitemOpen
  \bibfield  {author} {\bibinfo {author} {\bibfnamefont {S.}~\bibnamefont
  {Qiu}}, \bibinfo {author} {\bibfnamefont {X.}~\bibnamefont {Ma}}, \bibinfo
  {author} {\bibfnamefont {B.}~\bibnamefont {Huang}}, \bibinfo {author}
  {\bibfnamefont {D.}~\bibnamefont {Li}}, \bibinfo {author} {\bibfnamefont
  {G.}~\bibnamefont {Wang}}, \ and\ \bibinfo {author} {\bibfnamefont
  {M.}~\bibnamefont {Zhang}},\ }\bibfield  {title} {\enquote {\bibinfo {title}
  {Numerical simulation of single bubble dynamics under acoustic standing
  waves},}\ }\href@noop {} {\bibfield  {journal} {\bibinfo  {journal}
  {Ultrason. Sonochem.}\ }\textbf {\bibinfo {volume} {49}},\ \bibinfo {pages}
  {196--205} (\bibinfo {year} {2018})}\BibitemShut {NoStop}%
\bibitem [{\citenamefont {Boyd}\ and\ \citenamefont {Becker}(2019)}]{bb19}%
  \BibitemOpen
  \bibfield  {author} {\bibinfo {author} {\bibfnamefont {B.}~\bibnamefont
  {Boyd}}\ and\ \bibinfo {author} {\bibfnamefont {S.}~\bibnamefont {Becker}},\
  }\bibfield  {title} {\enquote {\bibinfo {title} {Numerical modeling of the
  acoustically driven growth and collapse of a cavitation bubble near a
  wall},}\ }\href@noop {} {\bibfield  {journal} {\bibinfo  {journal} {Phys.
  Fluids}\ }\textbf {\bibinfo {volume} {31}},\ \bibinfo {pages} {032102}
  (\bibinfo {year} {2019})}\BibitemShut {NoStop}%
\bibitem [{\citenamefont {Pandey}(2019)}]{pandey19}%
  \BibitemOpen
  \bibfield  {author} {\bibinfo {author} {\bibfnamefont {V.}~\bibnamefont
  {Pandey}},\ }\bibfield  {title} {\enquote {\bibinfo {title} {Asymmetricity
  and sign reversal of secondary bjerknes force from strong nonlinear coupling
  in cavitation bubble pairs},}\ }\href@noop {} {\bibfield  {journal} {\bibinfo
   {journal} {Phys. Rev. E}\ }\textbf {\bibinfo {volume} {99}},\ \bibinfo
  {pages} {042209} (\bibinfo {year} {2019})}\BibitemShut {NoStop}%
\bibitem [{\citenamefont {Yamamoto}\ and\ \citenamefont
  {Komarov}(2020)}]{yk20}%
  \BibitemOpen
  \bibfield  {author} {\bibinfo {author} {\bibfnamefont {T.}~\bibnamefont
  {Yamamoto}}\ and\ \bibinfo {author} {\bibfnamefont {S.~V.}\ \bibnamefont
  {Komarov}},\ }\bibfield  {title} {\enquote {\bibinfo {title} {Dynamic
  behavior of acoustic cavitation bubble originated from heterogeneous
  nucleation},}\ }\href@noop {} {\bibfield  {journal} {\bibinfo  {journal} {J.
  Appl. Phys.}\ }\textbf {\bibinfo {volume} {128}},\ \bibinfo {pages} {044702}
  (\bibinfo {year} {2020})}\BibitemShut {NoStop}%
\bibitem [{\citenamefont {Okumura}\ and\ \citenamefont {Ito}(2003)}]{oi03}%
  \BibitemOpen
  \bibfield  {author} {\bibinfo {author} {\bibfnamefont {H.}~\bibnamefont
  {Okumura}}\ and\ \bibinfo {author} {\bibfnamefont {N.}~\bibnamefont {Ito}},\
  }\bibfield  {title} {\enquote {\bibinfo {title} {Nonequilibrium molecular
  dynamics simulations of a bubble},}\ }\href@noop {} {\bibfield  {journal}
  {\bibinfo  {journal} {Phys. Rev. E}\ }\textbf {\bibinfo {volume} {67}},\
  \bibinfo {pages} {045301} (\bibinfo {year} {2003})}\BibitemShut {NoStop}%
\bibitem [{\citenamefont {Ho{\l}yst}, \citenamefont {Litniewski},\ and\
  \citenamefont {Garstecki}(2010)}]{hlg10}%
  \BibitemOpen
  \bibfield  {author} {\bibinfo {author} {\bibfnamefont {R.}~\bibnamefont
  {Ho{\l}yst}}, \bibinfo {author} {\bibfnamefont {M.}~\bibnamefont
  {Litniewski}}, \ and\ \bibinfo {author} {\bibfnamefont {P.}~\bibnamefont
  {Garstecki}},\ }\bibfield  {title} {\enquote {\bibinfo {title} {Large-scale
  molecular dynamics verification of the rayleigh-plesset approximation for
  collapse of nanobubbles},}\ }\href@noop {} {\bibfield  {journal} {\bibinfo
  {journal} {Phys. Revi. E}\ }\textbf {\bibinfo {volume} {82}},\ \bibinfo
  {pages} {066309} (\bibinfo {year} {2010})}\BibitemShut {NoStop}%
\bibitem [{\citenamefont {Watanabe}, \citenamefont {Ito},\ and\ \citenamefont
  {Hu}(2012)}]{wih12}%
  \BibitemOpen
  \bibfield  {author} {\bibinfo {author} {\bibfnamefont {H.}~\bibnamefont
  {Watanabe}}, \bibinfo {author} {\bibfnamefont {N.}~\bibnamefont {Ito}}, \
  and\ \bibinfo {author} {\bibfnamefont {C.-K.}\ \bibnamefont {Hu}},\
  }\bibfield  {title} {\enquote {\bibinfo {title} {Phase diagram and
  universality of the lennard-jones gas-liquid system},}\ }\href@noop {}
  {\bibfield  {journal} {\bibinfo  {journal} {J. Chem. Phys.}\ }\textbf
  {\bibinfo {volume} {136}},\ \bibinfo {pages} {204102} (\bibinfo {year}
  {2012})}\BibitemShut {NoStop}%
\bibitem [{\citenamefont {Asano}, \citenamefont {Watanabe},\ and\ \citenamefont
  {Noguchi}(2020{\natexlab{a}})}]{awn20b}%
  \BibitemOpen
  \bibfield  {author} {\bibinfo {author} {\bibfnamefont {Y.}~\bibnamefont
  {Asano}}, \bibinfo {author} {\bibfnamefont {H.}~\bibnamefont {Watanabe}}, \
  and\ \bibinfo {author} {\bibfnamefont {H.}~\bibnamefont {Noguchi}},\
  }\bibfield  {title} {\enquote {\bibinfo {title} {Molecular dynamics
  simulation of soundwave propagation in a simple fluid},}\ }\href@noop {}
  {\bibfield  {journal} {\bibinfo  {journal} {J. Chem. Phys.}\ }\textbf
  {\bibinfo {volume} {153}},\ \bibinfo {pages} {124504} (\bibinfo {year}
  {2020}{\natexlab{a}})}\BibitemShut {NoStop}%
\bibitem [{\citenamefont {Asano}, \citenamefont {Watanabe},\ and\ \citenamefont
  {Noguchi}(2018)}]{awn18}%
  \BibitemOpen
  \bibfield  {author} {\bibinfo {author} {\bibfnamefont {Y.}~\bibnamefont
  {Asano}}, \bibinfo {author} {\bibfnamefont {H.}~\bibnamefont {Watanabe}}, \
  and\ \bibinfo {author} {\bibfnamefont {H.}~\bibnamefont {Noguchi}},\
  }\bibfield  {title} {\enquote {\bibinfo {title} {Polymer effects on
  k\'arm\'an vortex: Molecular dynamics study},}\ }\href@noop {} {\bibfield
  {journal} {\bibinfo  {journal} {J. Chem. Phys.}\ }\textbf {\bibinfo {volume}
  {148}},\ \bibinfo {pages} {144901} (\bibinfo {year} {2018})}\BibitemShut
  {NoStop}%
\bibitem [{\citenamefont {Asano}, \citenamefont {Watanabe},\ and\ \citenamefont
  {Noguchi}(2019)}]{awn19}%
  \BibitemOpen
  \bibfield  {author} {\bibinfo {author} {\bibfnamefont {Y.}~\bibnamefont
  {Asano}}, \bibinfo {author} {\bibfnamefont {H.}~\bibnamefont {Watanabe}}, \
  and\ \bibinfo {author} {\bibfnamefont {H.}~\bibnamefont {Noguchi}},\
  }\bibfield  {title} {\enquote {\bibinfo {title} {Finite-size effects on
  k{\'a}rm{\'a}n vortex in molecular dynamics simulation},}\ }\href@noop {}
  {\bibfield  {journal} {\bibinfo  {journal} {J. Phys. Soc. Jpn.}\ }\textbf
  {\bibinfo {volume} {88}},\ \bibinfo {pages} {075003} (\bibinfo {year}
  {2019})}\BibitemShut {NoStop}%
\bibitem [{\citenamefont {Asano}, \citenamefont {Watanabe},\ and\ \citenamefont
  {Noguchi}(2020{\natexlab{b}})}]{awn20a}%
  \BibitemOpen
  \bibfield  {author} {\bibinfo {author} {\bibfnamefont {Y.}~\bibnamefont
  {Asano}}, \bibinfo {author} {\bibfnamefont {H.}~\bibnamefont {Watanabe}}, \
  and\ \bibinfo {author} {\bibfnamefont {H.}~\bibnamefont {Noguchi}},\
  }\bibfield  {title} {\enquote {\bibinfo {title} {Effects of cavitation on
  k\'arm\'an vortex behind circular-cylinder arrays: A molecular dynamics
  study},}\ }\href@noop {} {\bibfield  {journal} {\bibinfo  {journal} {J. Chem.
  Phys.}\ }\textbf {\bibinfo {volume} {152}},\ \bibinfo {pages} {034501}
  (\bibinfo {year} {2020}{\natexlab{b}})}\BibitemShut {NoStop}%
\bibitem [{\citenamefont {Asano}, \citenamefont {Watanabe},\ and\ \citenamefont
  {Noguchi}(2021)}]{awn21}%
  \BibitemOpen
  \bibfield  {author} {\bibinfo {author} {\bibfnamefont {Y.}~\bibnamefont
  {Asano}}, \bibinfo {author} {\bibfnamefont {H.}~\bibnamefont {Watanabe}}, \
  and\ \bibinfo {author} {\bibfnamefont {H.}~\bibnamefont {Noguchi}},\
  }\bibfield  {title} {\enquote {\bibinfo {title} {Effects of polymers on the
  cavitating flow around a cylinder: A large-scale molecular dynamics
  analysis},}\ }\href@noop {} {\bibfield  {journal} {\bibinfo  {journal} {J.
  Chem. Phys.}\ }\textbf {\bibinfo {volume} {155}},\ \bibinfo {pages} {014905}
  (\bibinfo {year} {2021})}\BibitemShut {NoStop}%
\bibitem [{\citenamefont {Plimpton}(1995)}]{plimpton95}%
  \BibitemOpen
  \bibfield  {author} {\bibinfo {author} {\bibfnamefont {S.}~\bibnamefont
  {Plimpton}},\ }\bibfield  {title} {\enquote {\bibinfo {title} {Fast parallel
  algorithms for short-range molecular dynamics},}\ }\href@noop {} {\bibfield
  {journal} {\bibinfo  {journal} {J. Comput. Phys.}\ }\textbf {\bibinfo
  {volume} {117}},\ \bibinfo {pages} {1--19} (\bibinfo {year}
  {1995})}\BibitemShut {NoStop}%
\bibitem [{\citenamefont {Burgers}(1948)}]{burgers48}%
  \BibitemOpen
  \bibfield  {author} {\bibinfo {author} {\bibfnamefont {J.~M.}\ \bibnamefont
  {Burgers}},\ }\bibfield  {title} {\enquote {\bibinfo {title} {A mathematical
  model illustrating the theory of turbulence},}\ \ }(\bibinfo  {publisher}
  {Elsevier},\ \bibinfo {year} {1948})\ pp.\ \bibinfo {pages}
  {171--199}\BibitemShut {NoStop}%
\bibitem [{\citenamefont {Mohamed}(2019)}]{mohamed19}%
  \BibitemOpen
  \bibfield  {author} {\bibinfo {author} {\bibfnamefont {N.~A.}\ \bibnamefont
  {Mohamed}},\ }\bibfield  {title} {\enquote {\bibinfo {title} {Solving one-and
  two-dimensional unsteady burgers' equation using fully implicit finite
  difference schemes},}\ }\href@noop {} {\bibfield  {journal} {\bibinfo
  {journal} {Arab J. Basic Appl. Sci.}\ }\textbf {\bibinfo {volume} {26}},\
  \bibinfo {pages} {254--268} (\bibinfo {year} {2019})}\BibitemShut {NoStop}%
\bibitem [{\citenamefont {Prosperetti}(2017)}]{prosperetti17}%
  \BibitemOpen
  \bibfield  {author} {\bibinfo {author} {\bibfnamefont {A.}~\bibnamefont
  {Prosperetti}},\ }\bibfield  {title} {\enquote {\bibinfo {title} {Vapor
  bubbles},}\ }\href@noop {} {\bibfield  {journal} {\bibinfo  {journal} {Annu.
  Rev. Fluid Mech.}\ }\textbf {\bibinfo {volume} {49}},\ \bibinfo {pages}
  {221--248} (\bibinfo {year} {2017})}\BibitemShut {NoStop}%
\bibitem [{\citenamefont {Park}\ \emph {et~al.}(2021)\citenamefont {Park},
  \citenamefont {Choi}, \citenamefont {Park}, \citenamefont {Shon},
  \citenamefont {Kim},\ and\ \citenamefont {Kim}}]{pcp21}%
  \BibitemOpen
  \bibfield  {author} {\bibinfo {author} {\bibfnamefont {R.}~\bibnamefont
  {Park}}, \bibinfo {author} {\bibfnamefont {M.}~\bibnamefont {Choi}}, \bibinfo
  {author} {\bibfnamefont {E.~H.}\ \bibnamefont {Park}}, \bibinfo {author}
  {\bibfnamefont {W.-J.}\ \bibnamefont {Shon}}, \bibinfo {author}
  {\bibfnamefont {H.-Y.}\ \bibnamefont {Kim}}, \ and\ \bibinfo {author}
  {\bibfnamefont {W.}~\bibnamefont {Kim}},\ }\bibfield  {title} {\enquote
  {\bibinfo {title} {Comparing cleaning effects of gas and vapor bubbles in
  ultrasonic fields},}\ }\href@noop {} {\bibfield  {journal} {\bibinfo
  {journal} {Ultrason. Sonochem.}\ ,\ \bibinfo {pages} {105618}} (\bibinfo
  {year} {2021})}\BibitemShut {NoStop}%
\bibitem [{\citenamefont {Tamidi}, \citenamefont {Lau},\ and\ \citenamefont
  {Khalit}(2021)}]{tlk21}%
  \BibitemOpen
  \bibfield  {author} {\bibinfo {author} {\bibfnamefont {A.}~\bibnamefont
  {Tamidi}}, \bibinfo {author} {\bibfnamefont {K.}~\bibnamefont {Lau}}, \ and\
  \bibinfo {author} {\bibfnamefont {S.}~\bibnamefont {Khalit}},\ }\bibfield
  {title} {\enquote {\bibinfo {title} {A review of recent development in
  numerical simulation of ultrasonic-assisted gas-liquid mass transfer
  process},}\ }\href@noop {} {\bibfield  {journal} {\bibinfo  {journal}
  {Comput. Chem. Eng.}\ ,\ \bibinfo {pages} {107498}} (\bibinfo {year}
  {2021})}\BibitemShut {NoStop}%
\bibitem [{\citenamefont {Lebon}\ \emph {et~al.}(2017)\citenamefont {Lebon},
  \citenamefont {Tzanakis}, \citenamefont {Djambazov}, \citenamefont
  {Pericleous},\ and\ \citenamefont {Eskin}}]{ltd17}%
  \BibitemOpen
  \bibfield  {author} {\bibinfo {author} {\bibfnamefont {G.~B.}\ \bibnamefont
  {Lebon}}, \bibinfo {author} {\bibfnamefont {I.}~\bibnamefont {Tzanakis}},
  \bibinfo {author} {\bibfnamefont {G.}~\bibnamefont {Djambazov}}, \bibinfo
  {author} {\bibfnamefont {K.}~\bibnamefont {Pericleous}}, \ and\ \bibinfo
  {author} {\bibfnamefont {D.}~\bibnamefont {Eskin}},\ }\bibfield  {title}
  {\enquote {\bibinfo {title} {Numerical modelling of ultrasonic waves in a
  bubbly newtonian liquid using a high-order acoustic cavitation model},}\
  }\href@noop {} {\bibfield  {journal} {\bibinfo  {journal} {Ultrason.
  Sonochem.}\ }\textbf {\bibinfo {volume} {37}},\ \bibinfo {pages} {660--668}
  (\bibinfo {year} {2017})}\BibitemShut {NoStop}%
\bibitem [{\citenamefont {Man}\ \emph {et~al.}(2018)\citenamefont {Man},
  \citenamefont {Li}, \citenamefont {Derreumaux},\ and\ \citenamefont
  {Nguyen}}]{mld18}%
  \BibitemOpen
  \bibfield  {author} {\bibinfo {author} {\bibfnamefont {V.~H.}\ \bibnamefont
  {Man}}, \bibinfo {author} {\bibfnamefont {M.~S.}\ \bibnamefont {Li}},
  \bibinfo {author} {\bibfnamefont {P.}~\bibnamefont {Derreumaux}}, \ and\
  \bibinfo {author} {\bibfnamefont {P.~H.}\ \bibnamefont {Nguyen}},\ }\bibfield
   {title} {\enquote {\bibinfo {title} {Rayleigh-plesset equation of the bubble
  stable cavitation in water: A nonequilibrium all-atom molecular dynamics
  simulation study},}\ }\href@noop {} {\bibfield  {journal} {\bibinfo
  {journal} {J. Chem. Phys.}\ }\textbf {\bibinfo {volume} {148}},\ \bibinfo
  {pages} {094505} (\bibinfo {year} {2018})}\BibitemShut {NoStop}%
\bibitem [{\citenamefont {Maeda}\ and\ \citenamefont {Maxwell}(2021)}]{mm21}%
  \BibitemOpen
  \bibfield  {author} {\bibinfo {author} {\bibfnamefont {K.}~\bibnamefont
  {Maeda}}\ and\ \bibinfo {author} {\bibfnamefont {A.~D.}\ \bibnamefont
  {Maxwell}},\ }\bibfield  {title} {\enquote {\bibinfo {title} {Controlling the
  dynamics of cloud cavitation bubbles through acoustic feedback},}\
  }\href@noop {} {\bibfield  {journal} {\bibinfo  {journal} {Phys. Rev. Appl.}\
  }\textbf {\bibinfo {volume} {15}},\ \bibinfo {pages} {034033} (\bibinfo
  {year} {2021})}\BibitemShut {NoStop}%
\end{thebibliography}
\providecommand{\noopsort}[1]{}\providecommand{\singleletter}[1]{#1}%

\end{document}